\documentclass[12pt]{article}
\usepackage[margin=0.5in]{geometry} 
\usepackage{latexsym}
\usepackage{amstext}
\usepackage{amsmath}
\usepackage[normalem]{ulem}
\usepackage{subfigure}
\usepackage{epsfig, graphicx}
\usepackage{color}
\usepackage{amsfonts,bm,color}
\usepackage{verbatim}
\usepackage{float}
\usepackage{latexsym}
\usepackage[svgnames]{xcolor}
\usepackage{colortbl}
\usepackage{float}
\usepackage{wrapfig}
\usepackage{multirow}
\usepackage{hyperref}
\usepackage{url}
\numberwithin{equation}{section}
\newcommand{\R}{\mathbb{R}}
\newcommand{\beq}{\begin{eqnarray}}
\newcommand{\eeq}{\end{eqnarray}}
\newcommand{\beqq}{\begin{eqnarray*}}
\newcommand{\eeqq}{\end{eqnarray*}}
\newcommand{\p}{\partial}

\newcommand{\x}{\mbox{\boldmath$x$}}
\newcommand{\X}{\mbox{\boldmath$X$}}

\newcommand{\y}{\mbox{\boldmath$y$}}

\newcommand{\n}{\mbox{\boldmath$n$}}

\begin{document}
\title{Narrow escape in composite domains forming heterogeneous networks}
\author{{Fr\'ed\'eric Paquin-Lefebvre$^{1}$, Kanishka Basnayake$^{1}$ and David Holcman$^{1,2}$}
\footnote{$^{1}$ Group of Applied Mathematics and Computational Biology, IBENS, Ecole Normale Sup\'erieure-PSL,75005 Paris, France. $^2$ DAMTP, University Of Cambridge, CB3 0WA, United Kingdom.}}
\date{\today}
\maketitle
\begin{abstract}
Cellular networks are often composed of thin tubules connecting much larger node compartments. These structures serve for active or diffusion transport of proteins. Examples are glial networks in the brain, the endoplasmic reticulum in cells or dendritic spines located on dendrites. In this latter case, a large ball forming the head is connected by a narrow passage. In all cases, how the transport of molecules, ions or proteins is regulated determines the time scale of chemical reactions or signal transduction. In the present study, based on modeling diffusion in three dimensions, we compute the mean time for a Brownian particle to reach a narrow target inside such a composite network made of tubules connected to spherical nodes. We derive asymptotic formulas by solving a mixed Neumann-Dirichlet boundary value problem with small Dirichlet part. We first consider the general case of a network domain organized in a 2-D lattice structure that consists of spherical ball compartments connected via narrow cylindrical passages. For a single target located on the boundary of one of the spherical domains, we derive a sparse linear system of equations for each Mean First Passage Time (MFPT) averaged over the different compartments. We then consider a composite domain  consisting of a spherical head-like domain connected to a large cylinder via another narrow cylindrical neck. For Brownian particles starting within the narrow neck, we derive asymptotic formulas for the MFPT to reach a target on the spherical head. When diffusing particles can be absorbed  upon hitting additional absorbing boundaries of the large cylinder, we derive asymptotic formulas for the probability and conditional MFPT to reach a target.  We compare these formulas with numerical solutions of the mixed boundary value problem and with Brownian simulations, allowing to explore the range of parameters. To conclude, the present analysis reveals that the mean arrival time, driven by diffusion in heterogeneous networks, is controlled by the sizes of the target and the narrow passages, as well as the size of the containers at each node.
\end{abstract}
\section{Introduction}
\begin{figure}[http!]
\centering
\includegraphics[width=\textwidth]{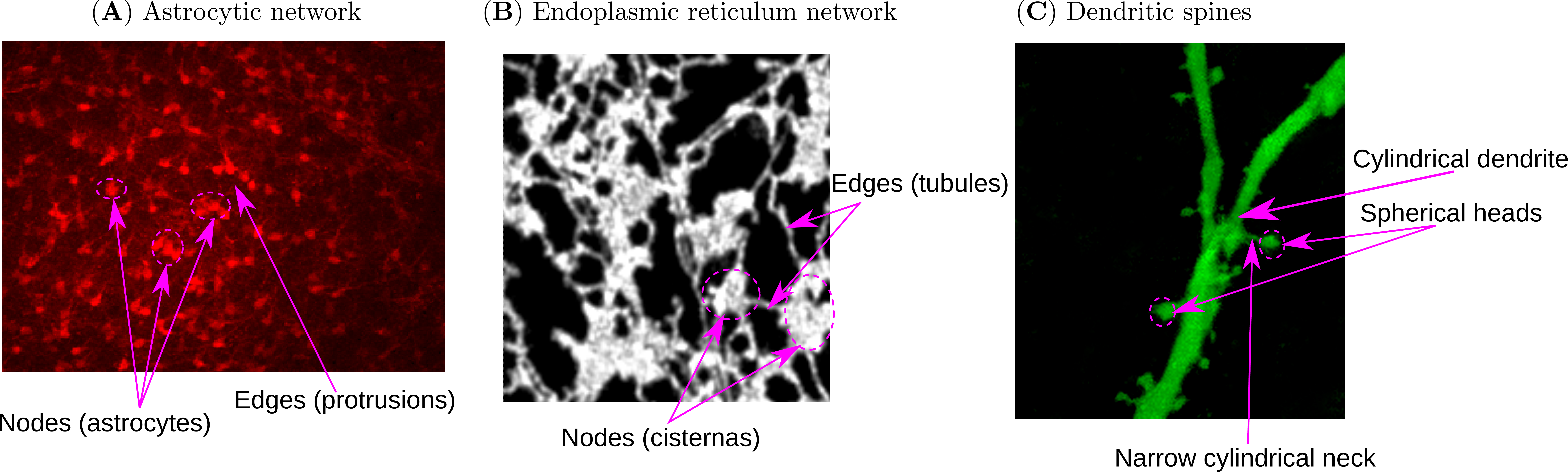}
\caption{\label{fig:fig1} \textbf{Biological networks illustrating composite domains.} \textbf{A}: Network of astrocytes in the brain is made of ball (red network) connected by narrow passage (courtesy of N. Rouach): a pipette delivers here a fluorescent dyes that diffuse passively inside the network. \textbf{B}: Endoplasmic reticulum network, where round nodes (cisterna) are connected by narrow cylinders  (courtesy of V. Kriechbaumer). \textbf{C}: Dendritic spines (white arrow) consist of a round head connecting a dendrite by a thin neck (courtesy of E. Korkotian).}
\end{figure}

\begin{figure}[http!]
\centering
\includegraphics[width=\textwidth]{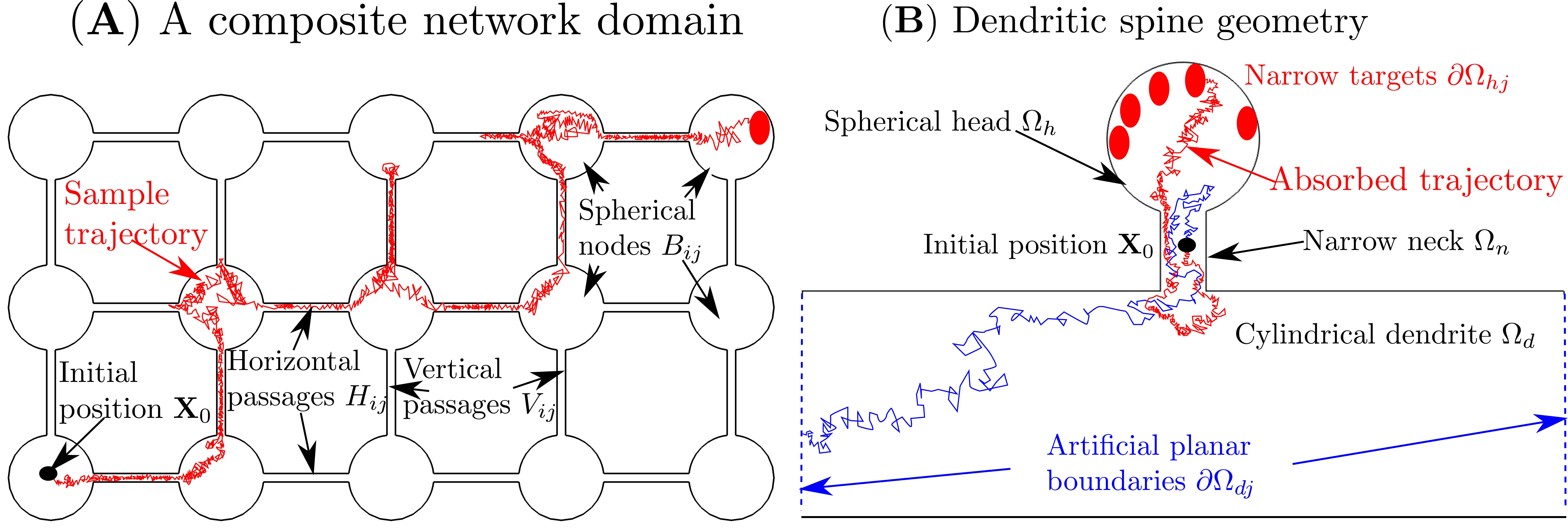}
\caption{\label{fig:fig2} \textbf{A}: Example of a network domain of balls connected by narrow cylinders, organized in a 2-D lattice. \textbf{B}: Idealized geometrical representation of a dendritic spine, composed of a spherical head connected to a large cylinder via a narrow cylindrical neck. Both panels show a 2-D frontview, and starting from a position $\bm{X}$ we give a few possible Brownian trajectories.}
\end{figure}

This manuscript describes a general approach to compute the mean time of a Brownian particle to reach a small target located inside a node of network made of narrow tubes connecting round balls (nodes). In that case, there is no possible reduction of the network three-dimensional geometry to a uniform narrow tube-shaped domains \cite{freidlin1996markov,rubinstein2001_I,rubinstein2001_II,rubinstein2001_III}, where the network structure converges, as the size of the tubule tends to zero, to a reduced one dimensional discrete graph embedded within the tube-shaped domains. We now start with some motivations arising from cell biology.\\
The hundreds of billions cells in the brain such as neurons, blood vessels or astrocytes \cite{kandel,alberts2013essential} are organized in interacting networks. Astrocytes are connected by tiny passages (connexin gap junction) \cite{rouach2000activity,rouach} allowing the passive diffusion of small particles (Fig.~\ref{fig:fig1}\textbf{A}) between the round cells. A key question remains to clarify the speed of potassium redistribution, calcium activation or energy recycling. Another example concerns the endoplasmic reticulum, a cellular organelle \cite{alberts2013essential}, where nanometer tubules (Fig.~\ref{fig:fig1}\textbf{B}) connect cisterna for
the transport and maturation of proteins circulating in the network. At cellular scale, we shall mention dendritic spines  (Fig.~\ref{fig:fig1}\textbf{C}), composed by spherical head connected to a narrow cylindrical neck \cite{yuste2010,basnayak2019Extreme,kushnireva2022} to the dendrite. The common denominators of these examples is that bulk compartments are connected by narrow passages. How these structures regulate molecular trafficking and diffusion remains unclear.\\
The context associated to computing asymptotic formula for the arrival time of a Brownian particle initially located at a point $\X$ inside a bounded domain $\Omega$ to a target (a narrow absorbing window $\p\Omega_A$ of radius $A$) is the narrow escape theory \cite{HolcmanPNAS2007,benichou2008,pillay2010,cheviakov2010,HolcmanSchuss2015,grebenkov2016,grebenkov2019}, where most of the boundary is a reflective surface. The Mean First Passage Time (MFPT) $T(\X)$, averaged over realizations is solution of the mixed boundary value problem \cite{schuss1980}
\beq\label{eq:intro_bvp}
\Delta T(\X) = -\frac{1}{D}\,, \quad \X \in \Omega\,,
\eeq
where $D$ is the diffusion coefficient of the underlying Brownian motion, with the boundary conditions
\beq\label{eq:intro_bc}
T(\X) = 0\,, \quad \X \in \p\Omega_A\,, \quad \frac{\p T(\X)}{\p\n} = 0\,, \quad \X \in \p\Omega \backslash \p\Omega_A\,.
\eeq
If the Dirichlet part is small enough compared to the boundary size, with $|\p\Omega_A| \ll |\p\Omega|$ and there are no smaller scale in the domain such as narrow passages \cite{HolcmanSchuss2015}, asymptotic analysis reveals that the leading order term of the expansion, outside a boundary layer near the absorbing window,
\beq\label{eq:intro}
T(\X) \sim \frac{|\Omega|}{4AD}\,,
\eeq
and thus the MFPT does not depend on the initial position, and for multiple well-spaced exits \eqref{eq:intro} is divided by the total number of windows. In fact the MFPT behaves as the reciprocal of the first eigenvalue of the Laplacian with mixed Neumann-Dirichlet boundary conditions, with such singularly perturbed eigenvalue problems studied in \cite{ward93,ward2005,davis2007,coombs2009,cheviakov2011}.\\
However the formula \eqref{eq:intro} ceases to be valid when the window is connected to the main bulk compartment via a narrow cylindrical neck (of radius $A$ as well), and rather than $T \sim O(1/A)$ the scaling law $T \sim O(1/A^2)$ is obtained for the MFPT \cite{biess2007diffusion,holcman2011diffusion}. This yields much longer escape times, despite the fact that diffusion within narrow passages essentially happens in 1-D and could be thought as facilitated. Brownian particles can indeed diffuse in and out of the narrow passage several times, thus making the event of finding the target rare.\\
In this article, we study the narrow escape theory on networks made of composite domains, where large 3-D compartments alternate with narrow almost one-dimensional structures. We will formulate the mixed boundary value problem \eqref{eq:intro_bvp}-\eqref{eq:intro_bc} on the two different domains illustrated in Fig.~\ref{fig:fig2}, and we shall derive asymptotic formulas for the MFPT highlighting the role played by narrow passages in controlling diffusion time scales. We shall focus on computing the diffusion time scales for such domains avoiding any singular limit as the radius of the narrow tubes tends to zero: the mixed boundary value problem \eqref{eq:intro_bvp}-\eqref{eq:intro_bc} is instead solved assuming radial symmetry within the narrow cylindrical passages, thereby yielding 1-D solutions.\\
We will consider first a general composite network domain where large spherical compartments are organized in a 2-D lattice structure, with narrow cylinders connecting each node and with a single absorbing target, as shown in Fig.~\ref{fig:fig2}\textbf{A}. For this case we will derive a sparse system of linear equations for the different MFPTs averaged over each compartment. Then we will consider a specific composite domain consisting of a spherical ball with multiple well-spaced absorbing targets, connected to a large cylindrical compartment via a narrow cylindrical passage as shown in Fig.~\ref{fig:fig2}\textbf{B}. For such a geometry inspired by the structure of the dendritic spine \cite{yuste2010}, we derive first explicit asymptotic formula for the MFPT assuming no loss from the large cylindrical bottom compartment (i.e.\ the opposite caps are reflecting). We then impose absorbing boundary conditions on the two flat boundaries of this large compartment, and compute the splitting probability to reach any targets located on the head boundary first, as well as the conditional MFPT.\\
The manuscript is organized as follows. We summarize after this paragraph the main asymptotic formulas derived in this manuscript. In Section \S \ref{sec:net}, we solve the narrow escape problem on a composite network domain with 2-D lattice structure. In Section \S \ref{sec:spine}, we consider the dendritic spine geometry, for which explicit asymptotic formulas for the MFPT are derived. Finally, we briefly summarize our results in Section \S \ref{sec:discussion} in the context of cellular biology. We also mention extensions that could warrant further investigation.
\section{Main asymptotic formulas derived in this manuscript}
\begin{figure}[http!]
\centering
\includegraphics[width=\textwidth]{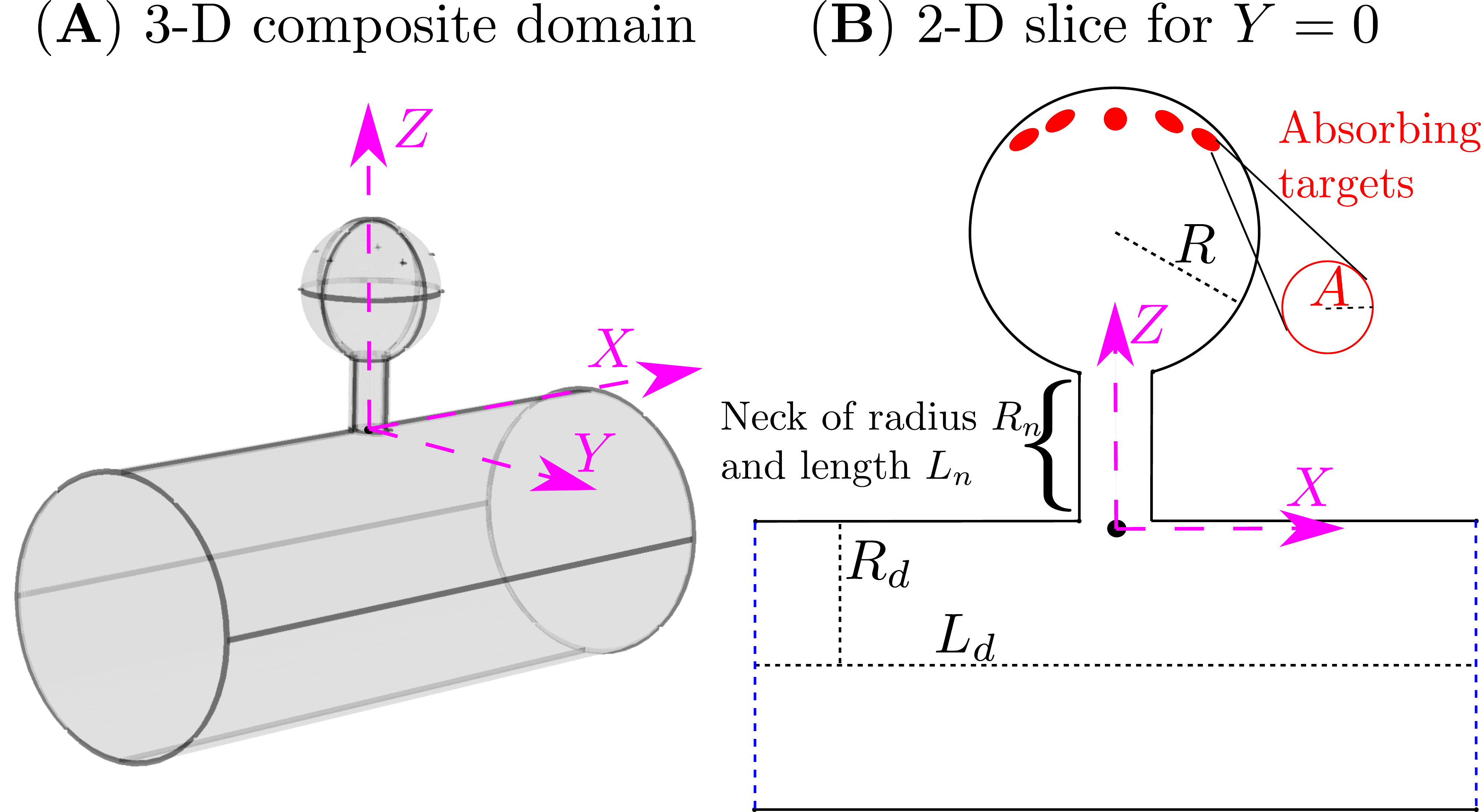}
\caption{\label{fig:fig3} \textbf{A}: Schematic representation of a dendritic spine geometry consisting of a spherical head connecting a dendritic with a cylindrical neck. \textbf{B}: 2-D frontview containing the main parameters of the model.}
\end{figure}
\subsection{Composite network domain}
We consider a network domain consisting of $M$ by $N$ 3-D balls of radius $R$ connected by narrow cylinders of radius $R_n$ and length $L_n$, with a single narrow absorbing window of radius $A$ on the last spherical compartment. For such a rectangular lattice structure, the main result of Section \S \ref{sec:net} is a matrix equation  for the vector $\bm{T}\in R^{MN}$ representing MFPTs averaged over each of the $MN$ network compartments:
\beq
\bm{T} = \left(\mathbb{D}_M \otimes I_N + I_M \otimes \mathbb{D}_N - \frac{4AL_n}{\pi R_n^2}\left(e_M^M\otimes e_N^N\right)\left(e_M^M \otimes e_N^N\right)^T\right)^{-1}\bm{\Xi}\,,
\eeq
with $\bm{\Xi} \in \R^{MN}$ given by
\beq
\begin{split}
\bm{\Xi} =& -\frac{4L_nR^3}{3R_n^2D}\left(\sum_{i=1}^{M}e_M^i\right) \otimes \left(\sum_{i=1}^{N}e_N^j\right)\\
&- \frac{L_n^2}{D}\left(\left(\frac{3}{2}\left(e_M^1 + e_M^M\right) + 2\sum_{i=2}^{M-1}e_M^i\right) \otimes \left(\sum_{j=1}^N e_N^j\right) - \frac{1}{2}\left(\sum_{i=1}^Me_M^i \right)\otimes\left( e_N^1 + e_N^N \right)\right)\,,
\end{split}
\eeq
where $D$ is the diffusion coefficient. Here the vectors $e_M^i$ for $i = 1,\ldots,M$ and $e_N^j$ for $j = 1, \ldots, N$ form the standard orthonormal Cartesian basis in $\R^M$ and $\R^N$ respectively, while $\mathbb{D}_{M} \in \R^{M\times M}$ and $\mathbb{D}_{N} \in \R^{N\times N}$ are tridiagonal matrices corresponding to the 1-D discrete Laplacian with reflecting boundary conditions, and $\otimes$ is the Kronecker product.
\subsection{Dendritic spine geometry}
For a composite domain $\Omega$ consisting of a ball of radius $R$ connected to a large cylinder of radius $R_d$ and length $L_d$ via a narrow cylindrical neck of radius $R_n$ and length $L_n$, and with $n_r$ narrow absorbing targets of radius $A$ on the head boundary (see Fig.~\ref{fig:fig3}), we find in Section \ref{sec:spine} the following asymptotic formula for the MFPT,
\beq
T_n(Z) \approx \frac{|\Omega|}{4Dn_rA} + \frac{L_dR_d^2}{DR_n}\left(1 + \frac{L_nR_n^2}{L_dR_d^2}\right)+ \frac{L_nL_dR_d^2}{DR_n^2}\left(1 + \frac{L_nR_n^2}{2L_dR_d^2}\right) - \frac{ZL_dR_d^2}{DR_n^2}\left(1 + \frac{ZR_n^2}{2L_dR_d^2}\right) \,,
\eeq
where the starting point $Z$, with $0 \leq Z \leq L_n$, measures the distance from the large cylindrical dendrite (Fig.~\ref{fig:fig3}) and $|\Omega| = \frac{4\pi R^3}{3} + \pi R_n^2L_n + \pi R_d^2L_d$ is the volume of the entire composite domain. Here $A$ and $R_n$ are two small geometrical parameters satisfying the inequality
\beq
\max\left\{A,R_n\right\} \ll \min\left\{R,R_d,L_n,L_d\right\}\,,
\eeq
and we refer to equation \eqref{eq:Lambda_0} for a refined approximation.\\
When the flat boundaries of the large cylindrical dendrite (see Fig.~\ref{fig:fig3}) are absorbing, we derive an expression for the splitting probability of trajectories that are reaching any targets before being absorbed within the cylindrical dendrite:
\beq
P_n(Z) \approx  \frac{1 + \frac{Z}{R_n}}{2 + \frac{L_n}{R_n} + \frac{\pi R_n}{4An_r}} \,, \quad \text{for } 0 \leq Z \leq L_n\,.
\eeq
In that case we also obtain the approximation below for the conditional MFPT
\beq
\begin{split}
T_n(Z) \approx& \frac{1}{2 + \frac{L_n}{R_n} + \frac{\pi R_n}{4An_r}}\left(\frac{\pi R^3}{3DAn_r}\left(2 + \frac{L_n}{R_n} + \frac{3R_nL_n^2}{8R^3} + \frac{3R_n^2L_n}{4R^3} \right) + \frac{L_n^2}{D}\left(1 + \frac{L_n}{6R_n} + \frac{R_n}{L_n}\right)\right) \\
&- \frac{Z^2}{6D}\left(1 + \frac{2}{1 + \frac{Z}{R_n}} \right)\,, \quad \text{for } 0 \leq Z \leq L_n\,.
\end{split}
\eeq
\section{Composite network domains}\label{sec:net}
\subsection{Definition of composite network domains}
\begin{figure}[http!]
\centering
\includegraphics[width=0.66\textwidth]{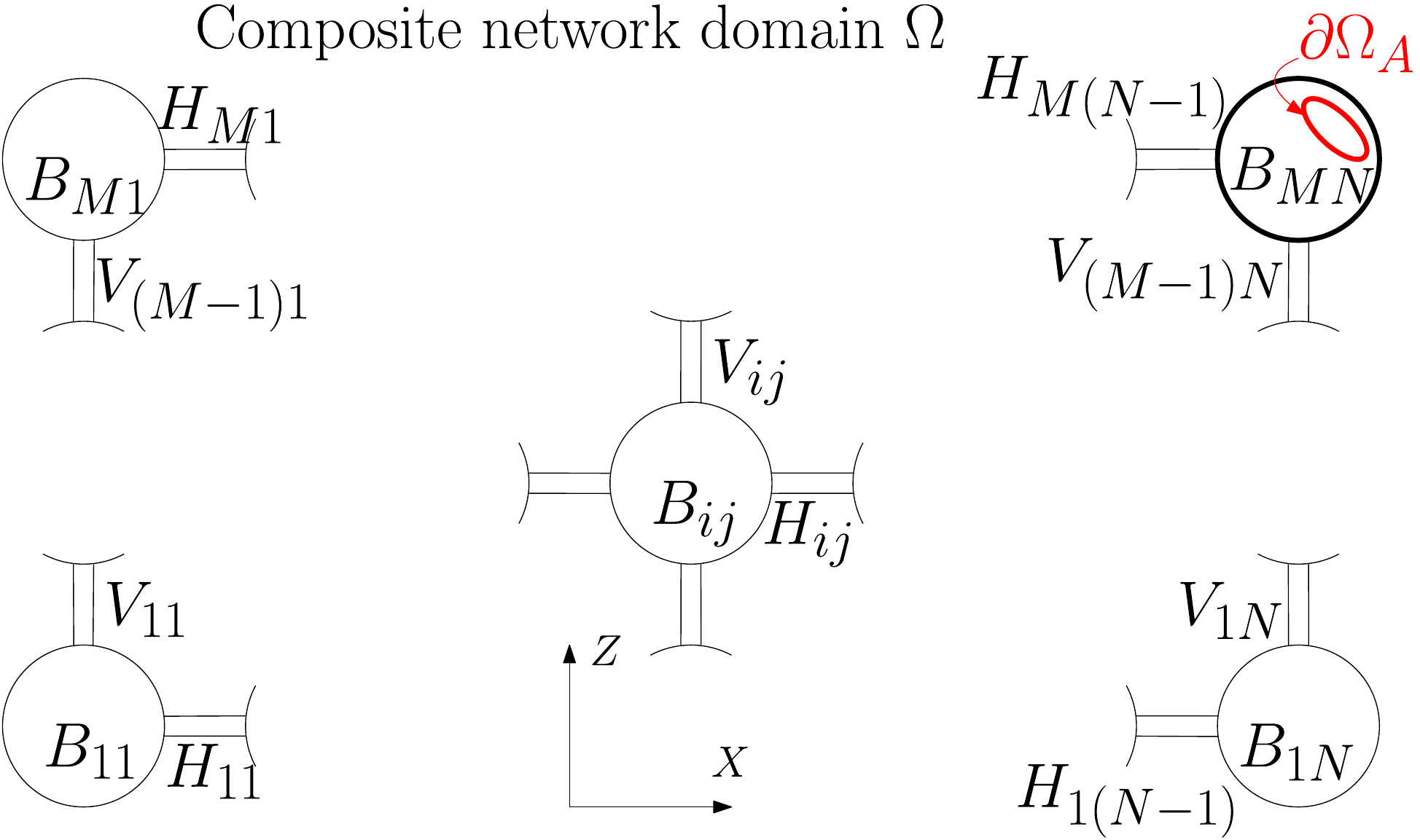}
\caption{\label{fig:network_2D_dim} \textbf{A rectangular lattice composite network domain.} Each ball of radius $R$ is connected to its nearest neighbors via four narrow cylindrical passages. The last lattice element $B_{MN}$ contains a narrow circular absorbing window $\p\Omega_A$ of radius $A$.}
\end{figure}

The composite network domain $\Omega$ made of a rectangular lattice with $M$ rows and $N$ columns is composed of 3-D balls connected by narrow cylindrical passages (Fig.~\ref{fig:network_2D_dim}). Each node is made of a spherical ball $B_{ij}$ of radius $R$ centered at the point $\X_{ij}$,
\beq
B_{ij} = \left\{ \left. \X \, \right| \, \|\X-\X_{ij}\| \leq R \right\}
\eeq
with $i=1,2,\ldots,M$ and $j=1,2,\ldots,N$. Horizontal narrow cylindrical neck passages $H_{ij}$ of length $L_n$ and radius $R_n$ connecting the nodes $B_{ij}$ and $B_{i(j+1)}$ are defined as
\beq
H_{ij} = \left\{ \X = (X,Y,Z) \, \left| \, \sqrt{Y^2 + (Z-Z_{ij})^2} \leq R_n,\, X_{ij} + R \leq X \leq X_{i(j+1)} - R \right. \right\}
\eeq
with $i=1,2,\ldots,M$ and $j=1,2,\ldots,N-1$, and vertical narrow cylindrical neck passages $V_{ij}$ connecting the nodes $B_{ij}$ and $B_{(i+1)j}$ are defined as
\beq
V_{ij} = \left\{ \X = (X,Y,Z) \, \left| \, \sqrt{Y^2 + (X-X_{ij})^2} \leq R_n,\, Z_{ij} + R \leq Z \leq Z_{(i+1)j} - R \right. \right\}
\eeq
with $i=1,2,\ldots,M-1$ and $j=1,2,\ldots,N$. \\
Using these definition, the composite domain $\Omega$ as
\beq
\Omega = \left(\cup_{i=1}^M\cup_{j=1}^N B_{ij}\right) \cup \left(\cup_{i=1}^M\cup_{j=1}^{N-1} H_{ij}\right) \cup \left(\cup_{i=1}^{M-1}\cup_{j=1}^N V_{ij}\right)\,.
\eeq
On the boundary of the last node $\p B_{MN}$, there is a narrow circular absorbing window $\p\Omega_A$ of radius $A$ centered in $\X_A$, such that the full boundary of the composite domain can be decomposed as
\beq
\p\Omega = \p\Omega_r\cup\p\Omega_A,
\eeq
where $\p\Omega_r$ is the reflecting part and $\p\Omega_A$ is absorbing. In the present manuscript, the absorbing windows are well-separated from the narrow cylindrical passages connecting to the rest of the network.
The dynamics inside the network is classical diffusion made of Brownian particles $W(\tau)$, with diffusion coefficient $D$ (no drift). The first escape time from the narrow absorbing target $\p\Omega_A$ when starting in $\X \in \Omega$ is
\beq
\tau_{\X} = \min\left\{ \tau \, | \, W(\tau) \in \p\Omega_A \mbox{ conditioned on } W(0) = \X \right\}\,.
\eeq
The Mean First Passage Time (MFPT) $T(\X)$ is defined as the average over several realizations $T(\X) = E(\tau_{\X})$, and it is solution of the Dynkin's mixed boundary value problem
\beq\label{eq:pde_network}
D \Delta T(\X) = -1\,, \quad \mbox{ for } \X \in \Omega\,,
\eeq
with boundary conditions
\beq\label{eq:bc_network}
\frac{\p T(\X)}{\p\n} = 0\,, \quad \mbox{ for } \X \in \p\Omega_r\,, \quad T(\X) = 0\,, \quad \mbox{ for } \X \in \p\Omega_A\,.
\eeq
In what follows, we shall derive a system of equations for the average MFPTs defined as
\beq
\overline{T_{ij}} = \frac{1}{|B_{ij}|}\int_{B_{ij}} T(\X) d\X\,,
\eeq
in the asymptotic limit of narrow cylindrical passages and absorbing window, such that
\beq
A \ll R,\, L_n \,, \quad \mbox{ and } \quad R_n \ll R,\, L_n\,.
\eeq
\subsection{Deriving a linear system of equations for the MFPT in a composite network}
Before deriving the linear system satisfied by the average MFPT, we proceed to a non-dimensionalization of \eqref{eq:pde_network} and \eqref{eq:bc_network} by using the common radius $R$ of the ball elements, assumed to be of the same order of the length of the cylindrical passages, i.\ e.\ with $R \sim O(L_n)$. We define the nondimensional variables
\beq
\x = (x,y,z) = \frac{\X}{R}\,, \quad t(\x) = \frac{D}{R^2}T(R\x)\,, \quad \mbox{ with } \x \in \omega = \frac{\Omega}{R}
\eeq
with the composite domain $\omega$ defined as the following union,
\beq
\omega = \left(\cup_{i=1}^M\cup_{j=1}^N b_{ij}\right) \cup \left(\cup_{i=1}^M\cup_{j=1}^{N-1} h_{ij}\right) \cup \left(\cup_{i=1}^{M-1}\cup_{j=1}^N v_{ij}\right)\,,
\eeq
where each $b_{ij}$ is a unit ball centered at the points $\x_{ij} = \X_{ij}/R$, while each $h_{ij}$ and $v_{ij}$ are narrow horizontal and vertical cylindrical passages of radius $r_n = R_n/R$ and length $l_n = L_n/R$. Finally the boundary $\p\omega$ is decomposed into a reflecting part $\p\omega_r$ and a narrow absorbing window $\p\omega_a$ centered in $\x_a = \X_a/R$ of radius $a = A/R$. Hence we seek to solve the dimensionless mixed boundary-value problem
\beq\label{eq:pde_network_ndim}
\Delta t(\x) = -1\,, \quad \mbox{ for } \x \in \omega\,,
\eeq
subject to
\beq\label{eq:bc_network_ndim}
\frac{\p t(\x)}{\p\n} = 0\,, \quad \mbox{ for } \x \in \p\omega_r\,, \quad t(\x) = 0\,, \quad \mbox{ for } \x \in \p\omega_a\,,
\eeq
by deriving a linear system of equations for the average MFPT $t_{ij}$ for each compartment,
\beq
\overline{t_{ij}} = \frac{1}{|b_{ij}|}\int_{b_{ij}}t(\x)d\x\,,
\eeq
for $i=1,\ldots,M$ and $j=1,\ldots,N$. We first proceed by reducing the dynamics through the narrow cylindrical passages to 1-D diffusion, yielding the approximation
\beq\label{eq:1D_diff}
t_{h_{ij}}(x) = t(\x)\,, \quad \mbox{ for } \x = (x,y,z) \in h_{ij}\,, \quad t_{v_{ij}}(z) = t(\x)\,, \quad \mbox{ for } \x = (x,y,z) \in v_{ij}\,.
\eeq
For each subdomain $b_{ij}$, we then introduce the Neumann-Green's function $G_s(\x,\y)$ solution of
\beq\label{eq:GF_pde_net}
\Delta G_s(\x;\y) = \frac{1}{|b_{ij}|}\,, \quad \mbox{ for } \x \in b_{ij}\,, \y \in \p b_{ij}\,, \quad \frac{\p}{\p\n}G_s(\x;\y) = \delta(\x-\y)\,, \quad \mbox{ for } \x\,, \y \in \p b_{ij}\,,
\eeq
and with $\int_{b_{ij}} G_s(\x;\y)d\x = 0$, which expands as
\beq\label{eq:GF_expan_net}
G_s(\x;\y) = \frac{1}{2\pi\|\x-\y\|} + O\left(\log\left(\|\x-\y\|\right)\right)\,, \quad 0 < \|\x-\y\| \ll 1\,,
\eeq
near the singular diagonal $\y=\x$. When $i \neq N$ and $j \neq M$ we apply Green's identity to \eqref{eq:pde_network_ndim}, \eqref{eq:bc_network_ndim} and \eqref{eq:GF_pde_net} and obtain
\beq
t_{h_{ij}}(x_{ij}+1) = \overline{t_{ij}} + \int_{\p b_{ij} \cap \p h_{ij}} G_s\left((\x;\x_{ij}+(1,0,0)^T\right) \frac{\p t(\x)}{\p\n} d\x + O(r_n^2)\,,
\eeq
which is readily evaluated using the integral computation from equation \eqref{eq:int2} of Appendix \ref{sec:head},
\beq
t_{h_{ij}}(x_{ij}+1) = \overline{t_{ij}} + r_n\left.\frac{d}{dx}t_{h_{ij}}(x)\right|_{x=x_{ij}+1} + O(r_n^2\log(r_n)) + O(r_n^2)\,,
\eeq
and finally by only keeping the average $\overline{t_{ij}}$ and dropping higher-order terms we get
\beq\label{eq:bdy1}
t_{h_{ij}}(x_{ij}+1) = \overline{t_{ij}} + O(r_n)\,.
\eeq
Similarly, we obtain for the other directions,
\beq\label{eq:bdy2}
t_{h_{ij}}(x_{i(j+1)}-1) = \overline{t_{i(j+1)}} + O(r_n)\,, \quad t_{v_{ij}}(z_{ij}+1) = \overline{t_{ij}} + O(r_n)\,, \quad t_{v_{ij}}(z_{(i+1)j}-1) = \overline{t_{(i+1)j}} + O(r_n)\,.
\eeq
In the one dimensional limit in \eqref{eq:1D_diff}, along with the boundary conditions given in \eqref{eq:bdy1} and \eqref{eq:bdy2} from which the $O(r_n)$ terms were neglected, we get the boundary value problems below,
\beq\label{eq:bvp_h}
\frac{d^2t_{h_{ij}}(x)}{dx^2} = -1\,, \quad \mbox{ for } x_{ij} + 1 \leq x \leq x_{i(j+1)} - 1\,, \quad t_{h_{ij}}(x_{ij}+1) = \overline{t_{ij}}, \quad t_{h_{ij}}(x_{i(j+1)}-1) = \overline{t_{i(j+1)}}\,,
\eeq
for $j=1,\ldots,N-1$, and
\beq\label{eq:bvp_v}
\frac{d^2t_{v_{ij}}(z)}{dz^2} = -1\,, \quad \mbox{ for } z_{ij} + 1 \leq z \leq z_{(i+1)j} - 1\,, \quad t_{v_{ij}}(z_{ij}+1) = \overline{t_{ij}}, \quad t_{v_{ij}}(z_{(i+1)j}-1) = \overline{t_{(i+1)j}}\,.
\eeq
for $i=1,\ldots,N-1$. The solutions to \eqref{eq:bvp_h} and \eqref{eq:bvp_v} are
\beq
\begin{split}
&t_{h_{ij}}(x) = \\
&-\frac{x^2}{2} + x\left(\frac{\overline{t_{i(j+1)}} - \overline{t_{ij}}}{l_n} + \frac{x_{ij}+x_{i(j+1)}}{2} \right) + \frac{\overline{t_{ij}}(x_{i(j+1)}-1) - \overline{t_{i(j+1)}}(x_{ij}+1)}{l_n} - \frac{(x_{ij} + 1)(x_{i(j+1)} - 1)}{2}\,,
\end{split}
\eeq
and
\beq
\begin{split}
&t_{v_{ij}}(z) = \\
&-\frac{z^2}{2} + z\left(\frac{\overline{t_{(i+1)j}} - \overline{t_{ij}}}{l_n} + \frac{z_{ij}+z_{(i+1)j}}{2} \right) + \frac{\overline{t_{ij}}(z_{(i+1)j}-1) - \overline{t_{(i+1)j}}(z_{ij}+1)}{l_n} - \frac{(z_{ij} + 1)(z_{(i+1)j} - 1)}{2}\,.
\end{split}
\eeq
One condition to connect the various constant is to use the divergence theorem over each subdomain $b_{ij}$ for $i\neq 1,M$ and $j\neq 1,N$ which yields
\beq
\pi r_n^2 \left(t_{h_{ij}}'(x_{ij}+1) - t_{h_{i(j-1)}}'(x_{ij}-1) + t_{v_{ij}}'(z_{ij}+1) - t_{v_{(i-1)j}}'(z_{ij}-1)\right) = -|b_{ij}|\,,
\eeq
which becomes
\beq
\overline{t_{i(j+1)}} + \overline{t_{i(j-1)}} + \overline{t_{(i+1)j}} + \overline{t_{(i-1)j}} - 4\overline{t_{ij}} = -\frac{|b_{ij}|l_n}{\pi r_n^2} - 2l_n^2, \quad \mbox{ for } i = 2,\ldots,M-1,\, j = 2,\ldots,N-1\,.
\eeq
Similarly near the edges of the network we get the equations below
\begin{align}
\overline{t_{i2}} + \overline{t_{(i+1)1}} + \overline{t_{(i-1)1}} - 3\overline{t_{i1}} &=  -\frac{|b_{i1}|l_n}{\pi r_n^2} - \frac{3}{2}l_n^2\,, \quad i = 2,\ldots,M-1, \\
\overline{t_{i(N-1)}} + \overline{t_{(i+1)N}} + \overline{t_{(i-1)N}} - 3\overline{t_{iN}} &=  -\frac{|b_{iN}|l_n}{\pi r_n^2} - \frac{3}{2}l_n^2\,, \quad i = 2,\ldots,M-1, \\
\overline{t_{1(j+1)}} + \overline{t_{1(j-1)}} + \overline{t_{2j}} - 3\overline{t_{1j}} &=  -\frac{|b_{1j}|l_n}{\pi r_n^2} - \frac{3}{2}l_n^2\,, \quad j = 2,\ldots,N-1, \\
\overline{t_{M(j+1)}} + \overline{t_{M(j-1)}} + \overline{t_{(M-1)j}} - 3\overline{t_{Mj}} &=  -\frac{|b_{Mj}|l_n}{\pi r_n^2} - \frac{3}{2}l_n^2\,, \quad j = 2,\ldots,N-1,
\end{align}
while near each corner of the network, we obtain
\begin{align}
\overline{t_{21}} + \overline{t_{12}} - 2\overline{t_{11}} &=  -\frac{|b_{11}|l_n}{\pi r_n^2} - l_n^2\,, \\
\frac{\pi r_n^2}{l_n}\left(\overline{t_{(M-1)1}} + \overline{t_{M2}} - 2\overline{t_{M1}}\right) &=  -\frac{|b_{M1}|l_n}{\pi r_n^2} - l_n^2\,, \\
\overline{t_{2N}} + \overline{t_{1(N-1)}} - 2\overline{t_{1N}} &=  -\frac{|b_{1N}|l_n}{\pi r_n^2} - l_n^2\,.
\end{align}
Finally, when $i=M$ and $j=N$, we use the classical Weber's solution \cite{crank1975} to approximate the flux out of the narrow absorbing window $\p\omega_a$,
\beq
\frac{\p}{\p \n}t(\x) = \frac{C}{\sqrt{a^2 - \|\x-\x_a\|^2}}\, \quad \textbf{ for }\x \in \p\omega_a\,,
\eeq
where $C$ is an unknown constant, that we determine using Green's identity evaluated at the center of the exit window $\x_a$:
\beq
\begin{split}
t(\x_a) &= \overline{t_{MN}} + C\int_{\p \omega_{a}}\frac{G_s(\x;\y)}{\sqrt{a^2 - \|\x-\x_a\|^2}}d\x + O(r_n^2)\,.
\end{split}
\eeq
Using the identity $t(\x_a)=0$ as well as result \eqref{eq:int1} from Appendix \ref{sec:head}, we get the asymptotic relation,
\beq
0 = \overline{t_{MN}} + C \frac{\pi}{2} + O(a\log(a)) + O(r_n^2)\,.
\eeq
and thus
\beq\label{eq:C_flux}
C = -\frac{2}{\pi}\overline{t_{MN}} + O(a\log(a)) + O(r_n^2)\,.
\eeq
Finally, upon applying the divergence theorem over the domain $b_{MN}$ we get
\beq\label{eq:eq_N}
-\pi r_n^2 \left( t_{v_{(M-1)N}}'(z_{MN} - 1) + t_{h_{M(N-1)}}'(x_{MN} - 1) \right) + 2\pi a C = -|b_{MN}|\,,
\eeq
which becomes
\beq
\overline{t_{(M-1)N}} + \overline{t_{M(N-1)}} - \left(2 + \frac{4al_n}{\pi r_n^2}\right)\overline{t_{MN}} = -\frac{|b_{MN}|l_n}{\pi r_n^2} - l_n^2\,,
\eeq
after substituting the relation \eqref{eq:C_flux} and neglecting higher-order terms. We then define the vector $\bm{t} \in \R^{MN}$ of unknowns as
\beq
\bm{t} = \left( \overline{t_{11}},\,\ldots,\,\overline{t_{M1}},\,\overline{t_{12}},\,\ldots,\,\overline{t_{ij}},\,\overline{t_{(i+1)j}},\,\ldots,\,\overline{t_{M(N-1)}},\,\overline{t_{1N}},\,\ldots,\, \overline{t_{MN}} \right)^T\,,
\eeq
and derive the following sparse system of linear equations,
\beq\label{eq:sys2D}
\left(\mathbb{D}_M \otimes I_N + I_M \otimes \mathbb{D}_N - \frac{4al_n}{\pi r_n^2}\left(e_M^M\otimes e_N^N\right)\left(e_M^M \otimes e_N^N\right)^T\right)\bm{t} = \bm{\xi}\,.
\eeq
Since $|b_{ij}|=4\pi/3$ the right-hand side $\bm{\xi}$ can be written as
\beq
\begin{split}
\bm{\xi} =& -\frac{4l_n}{3r_n^2}\left(\sum_{i=1}^{M}e_M^i\right) \otimes \left(\sum_{i=1}^{N}e_N^j\right) \\
&- l_n^2\left(\left(\frac{3}{2}\left(e_M^1 + e_M^M\right) + 2\sum_{i=2}^{M-1}e_M^i\right) \otimes \left(\sum_{j=1}^N e_N^j\right) - \frac{1}{2}\left(\sum_{i=1}^Me_M^i \right)\otimes\left( e_N^1 + e_N^N \right)\right)\,,
\end{split}
\eeq
where $\otimes$ is the Kronecker product, $I_M \in \R^{M\times M}$ and $I_N \in \R^{N\times N}$ are identity matrices, $e_M^i \in \R^M$ for $i=1,\,2,\,\ldots,\,M$ and $e_N^j \in \R^N$ for $j=1,\,2,\,\ldots,\,N$ are the standard cartesian basis vectors. Moreover, $\mathbb{D}_M$ is a $M$ by $M$ tridiagonal matrix corresponding to the discrete one-dimensional Laplacian and defined as
\beq
\mathbb{D}_M =
\begin{pmatrix}
-1 & 1 & 0 & \cdots & 0 & 0 & 0\\
1 & -2 & 1 & \cdots & 0 & 0 & 0\\
\vdots & \vdots & \vdots & \ddots & \vdots & \vdots & \vdots \\
0 & 0 & 0 & \cdots & 1 & -2 & 1 \\
0 & 0 & 0 & \cdots & 0 & 1 & -1
\end{pmatrix} \, \in \R^{M \times M}\,,
\eeq
and similarly for $\mathbb{D}_N \in \R^{N \times N}$.\\
When the normalized radius of the absorbing window is zero $a = 0$, the linear system \eqref{eq:sys2D} becomes singular: on the left-hand side we get the two-dimensional discrete Laplacian operator with reflecting boundary conditions, whose solution is defined up to a constant. Hence we have derived a discrete version of the mixed Neumann-Dirichlet elliptic PDE problem for the MFPT time on a composite network domain with a 2-D lattice structure. \\
Finally to account for multiple exit sites, i.e.\ with an absorbing target well-separated from the narrow passages on each spherical node, we simply need to replace the rank-one perturbation matrix from \eqref{eq:sys2D} by the identity matrix to obtain
\beq
\left(\mathbb{D}_M \otimes I_N + I_M \otimes \mathbb{D}_N - \frac{4al_n}{\pi r_n^2}I_M \otimes I_N\right)\bm{t} = \bm{\xi}\,.
\eeq
\subsection{One-dimensional lattice structure}
\begin{figure}[http!]
\centering
\includegraphics[width=0.66\textwidth]{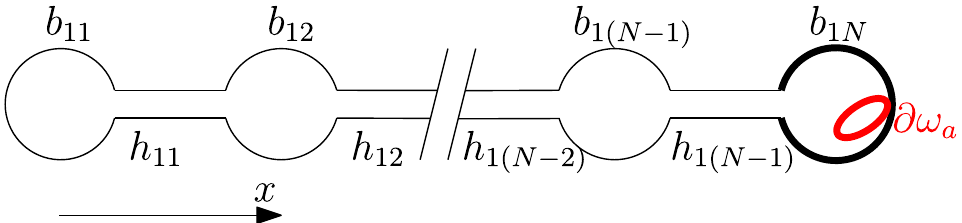}
\caption{\label{fig:network_1D_ndim} A composite network domain with one-dimensional lattice structure, composed of $N$ unit balls and $N-1$ narrow cylindrical passages each of length $l_n$.}
\end{figure}

When $M=1$, the lattice structure becomes one-dimensional (Fig.~\ref{fig:network_1D_ndim}) and using the identities $\mathbb{D}_1 = 0$, $I_1 = 1$ and $e_1^1 = 1$, the linear system \eqref{eq:sys2D} simplifies to
\beq\label{eq:sys1D}
\left(\mathbb{D}_N - \frac{4al_n}{\pi r_n^2}e_N^N\left(e_N^N\right)^T\right)\bm{t} = -\frac{4l_n}{3r_n^2}\left(\sum_{i=1}^N e_N^j\right) - l_n^2\left(\frac{1}{2}e_N^1 + \sum_{j=2}^{N-1} e_N^j + \frac{1}{2}e_N^N\right)
\,.
\eeq
\subsection{Numerical solutions of a mixed boundary-value problem for the MFPT in a network}
\begin{figure}[http!]
\centering
\includegraphics[width=0.66\textwidth]{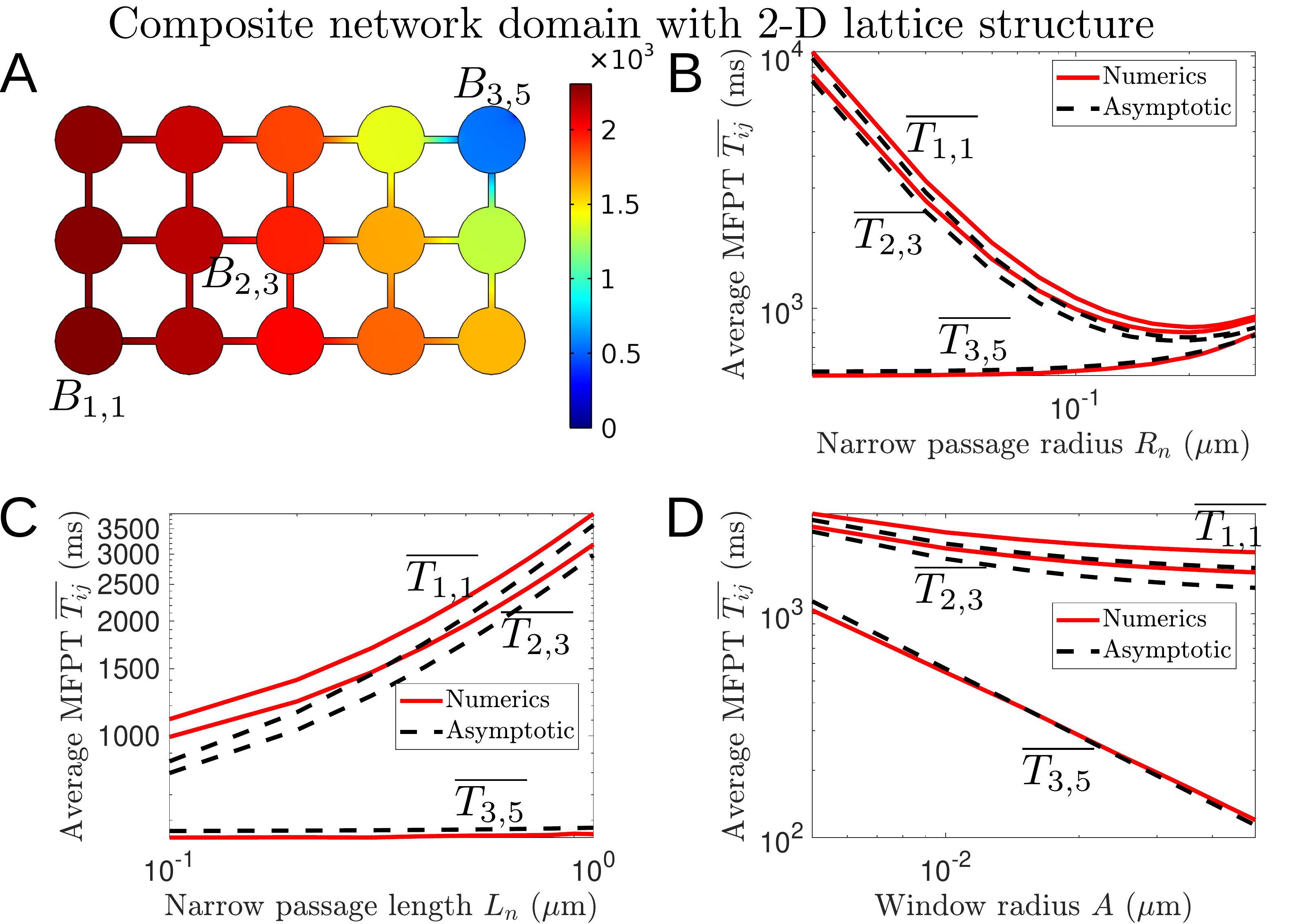}
\caption{\label{fig:TWOD} \textbf{Composite network domain with 2-D lattice structure}. \textbf{A}: 2-D frontview of the MFPT $T(\X)$ computed numerically by solving \eqref{eq:pde_network} and \eqref{eq:bc_network} with COMSOL, for a network with $3\times 5$ compartments. Parameter values are $D=0.35 \, \mu{\rm m}^2/{\rm ms}$, $R=0.5\,\mu{\rm m}$, $L_n=0.5\,\mu{\rm m}$, $R_n=0.05\,\mu{\rm m}$ and $A=0.01\,\mu{\rm m}$. The remaining panels show the average MFPT within the compartments $B_{1,1}$, $B_{2,3}$ and $B_{3,5}$ as a function of the radius $R_n$ (\textbf{B}) and length $L_n$ (\textbf{C}) of the narrow passages, and the target radius $A$ (\textbf{D}).}
\end{figure}

\begin{figure}[http!]
\centering
\includegraphics[width=0.66\textwidth]{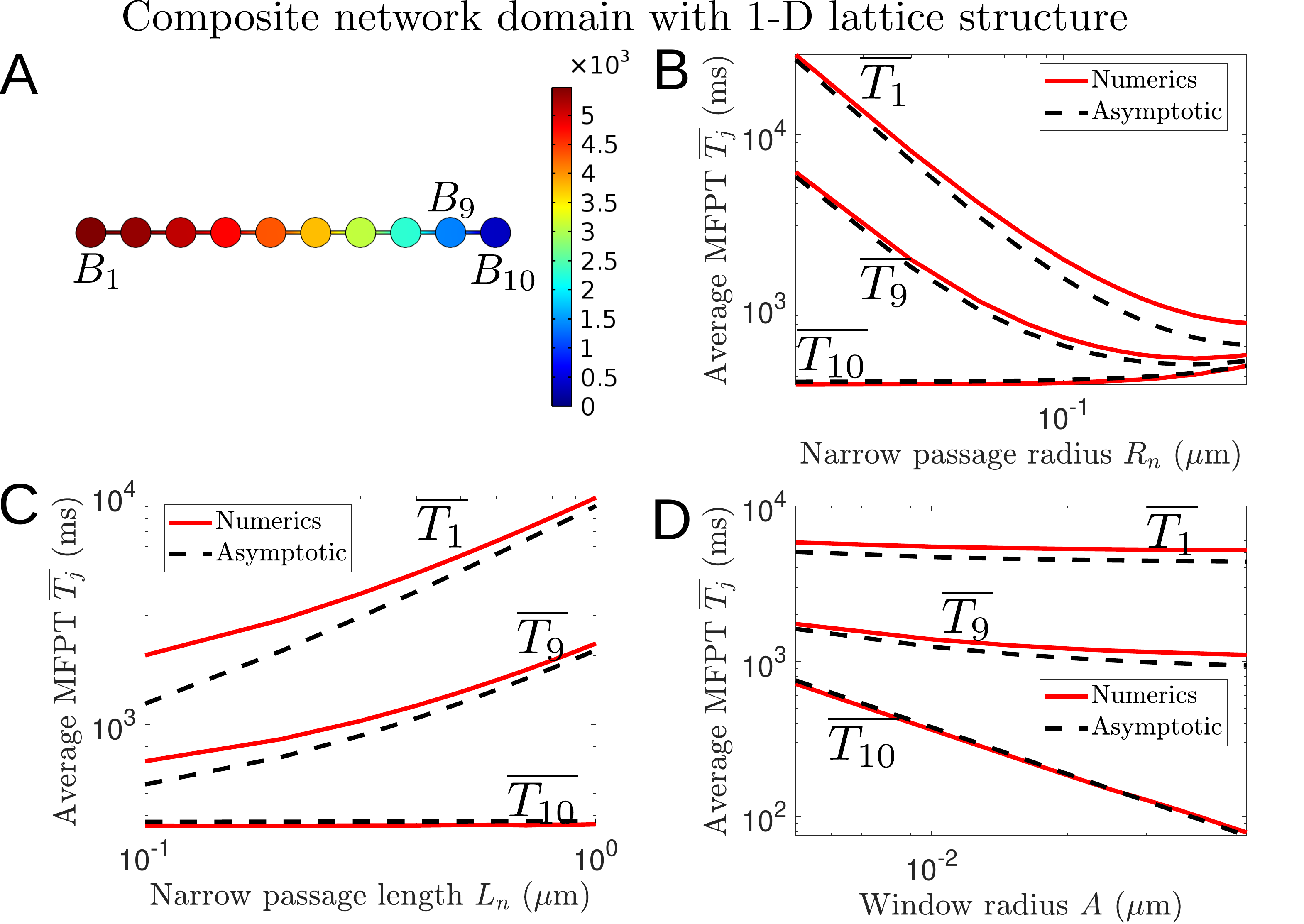}
\caption{\label{fig:ONED} \textbf{Composite network domain with 1-D lattice structure}. \textbf{A}: 2-D frontview of the MFPT $T(\X)$ computed numerically by solving \eqref{eq:pde_network} and \eqref{eq:bc_network} with COMSOL, for a network with $1\times 10$ compartments. Parameter values are $D=0.35 \, \mu{\rm m}^2/{\rm ms}$, $R=0.5\,\mu{\rm m}$, $L_n=0.5\,\mu{\rm m}$, $R_n=0.05\,\mu{\rm m}$ and $A=0.01\,\mu{\rm m}$. The remaining panels show the average MFPT within the compartments $B_1$, $B_9$ and $B_{10}$ as a function of the radius $R_n$ (\textbf{B}) and length $L_n$ (\textbf{C}) of the narrow passages, and the target radius $A$ (\textbf{D}).}
\end{figure}

In this section, we compare solutions of the linear systems \eqref{eq:sys2D} and \eqref{eq:sys1D} to direct numerical simulations of the mixed boundary-value problem \eqref{eq:pde_network_ndim} and \eqref{eq:bc_network_ndim} performed with COMSOL \cite{comsol}. Two different network domains are considered, one consisting of $M=3$ rows with $N=5$ columns (Fig.~\ref{fig:TWOD}) and the other of a chain of $N=10$ ball compartments (Fig.~\ref{fig:ONED}). The results, given in SI units, are comparable for both cases.\\
The average escape times are affected by geometrical parameters of the network, as illustrated in Fig.~\ref{fig:TWOD}\textbf{B}-\textbf{C} and Fig.~\ref{fig:ONED}\textbf{B}-\textbf{C} where we vary the radius and the length of the narrow passages. A good agreement is obtained between the numerical and asymptotic solutions, although a discrepancy is observed in Fig.~\ref{fig:ONED}\textbf{C} and Fig.~\ref{fig:TWOD}\textbf{C} when the networks are densely packed, i.e.\ with the length $L_n$ being small. Indeed for this parameter range we have $R_n \sim O(L_n)$ and the cylindrical passages are no longer narrow, explaining why the asymptotic theory ceases to be valid. Furthermore we find in Fig.~\ref{fig:ONED}\textbf{B} and Fig.~\ref{fig:TWOD}\textbf{B} that the average MFPT $\overline{T_{ij}}$ decreases with the narrow passage radius $R_n$, with the exception of the compartment containing the narrow exit window $\p\Omega_A$. Indeed, for Brownian particles starting within $B_{MN}$, bigger radius $R_n$ leads to longer search times since they can escape more easily through the narrow passages instead of directly reaching the absorbing target. Interestingly when the radius $R_n$ is large the different curves for each average MFPTs merge, indicating little influence from the location of the initial compartment.
\section{Time scale of diffusion between a dendrite and a dendritic spine}\label{sec:spine}
\begin{figure}[http!]
\centering
\includegraphics[width=0.5\textwidth]{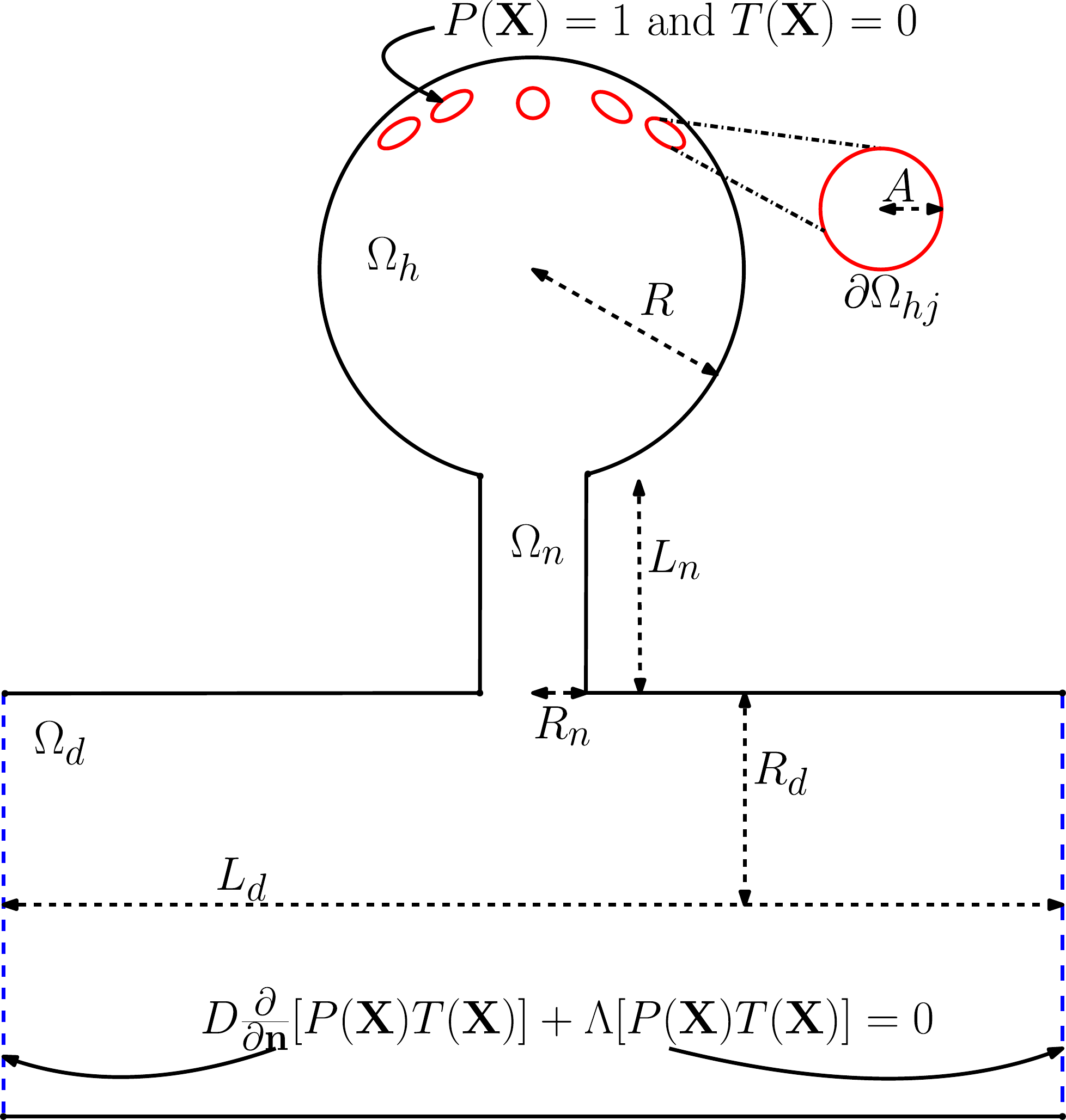}
\caption{\label{fig:mito_new} \textbf{Schematic diagram of a spine connected to a cylindrical dendrite.} The composite domain $\Omega$ consists of a spherical ball of radius $R$ on top of a narrow cylindrical passage of radius $R_n$ and length $L_n$, whose bottom end is connected to a larger cylindrical compartment of radius $R_d$ and length $L_d$.}
\end{figure}

In this section we study a specific 3-D composite domain inspired by the structure of the dendritic spines located on neuronal cells (Figs. \ref{fig:fig2} and \ref{fig:fig3}), for which we propose to derive explicit asymptotic solutions for the mean first passage times.\\
The domain $\Omega$ is divided into three compartments (Fig.~\ref{fig:mito_new}),
\beq
\Omega = \Omega_h \cup \Omega_n \cup \Omega_d,
\eeq
where the spherical head compartment $\Omega_h$ has radius $R$ and is parametrized  as
\beq
\Omega_h = \left\{ \X = (X,Y,Z) \, \left| \, X^2 + Y^2 + \left(Z - L_n - R \right)^2 \leq R^2 \right. \right\}.
\eeq
The cylindrical dendrite $\Omega_d$ has a radius $R_d$ and length $L_d$. Finally, $\Omega_n$ corresponding to the narrow cylindrical neck of radius $R_n$ and length $L_n$ connecting the head to the dendrite. The two cylinders are parametrized as
\begin{align}
\Omega_n &= \left\{ \X = (X,Y,Z) \, \left| \, X^2 + Y^2 \leq R_n^2 \text{ and } 0 \leq Z \leq L_n \right. \right\}\,, \\
\Omega_d &= \left\{ \X = (X,Y,Z) \, \left| \, Y^2 + (Z+R_d)^2 \leq R_d^2 \text{ and } -\frac{L_d}{2} \leq X \leq \frac{L_d}{2} \right. \right\}\,,
\end{align}
where the origin $\X = (0,0,0)$ is conveniently located at the bottom of the cylindrical neck. Since the cylindrical neck $\Omega_n$ acts as a narrow passage between the dendrite $\Omega_d$ and the spherical head $\Omega_h$, we imposes the conditions
\beq
\frac{R_n}{R} \ll 1\,, \quad \text{and} \quad \frac{R_n}{R_d} \ll 1.
\eeq
The entrance and the exit of the narrow passage are defined as
\beq
\begin{split}
\p\Omega_n\cap\p\Omega_d &= \left\{ \X = (X,Y,Z) \, \left| \, Z = 0 \text{ and } X^2 + Y^2 \leq R_n^2 \right. \right\}\,, \\
\p\Omega_n\cap\p\Omega_h &= \left\{ \X = (X,Y,Z) \, \left| \, Z = L_n \text{ and } X^2 + Y^2 \leq R_n^2 \right. \right\}\,,
\end{split}
\eeq
The boundary $\p\Omega$ is everywhere reflective except for a collection of $n_r$ identical narrow targets $\p\Omega_{hj}$ of radius $A$, located on the spherical boundary and thus defined as
\beq
\p\Omega_{hj} = \left\{ \X \in \p\Omega_h \, | \, \|\X - \X_j\| \leq A \right\}\,, \quad \text{for } j = 1,\ldots,n_r\,, \text{ with } \frac{A}{R} \ll 1.
\eeq
Brownian particles can escape either from the two planar boundaries of the dendritic cylinder $\Omega_d$ (blue line in Fig. \ref{fig:mito_new}) or from  circular planar absorbing disks, parametrized by
\beq
\p\Omega_{dj} = \left\{ \X = (X,Y,Z) \, \left| \, X = (-1)^j\frac{L_d}{2}\,, \quad Y^2 + (Z+R_d)^2 \leq R_d^2 \right.  \right\}\,, \quad j = 1,2\,.
\eeq
For a starting point $\X$ within the domain $\Omega$, we recall the definition of the splitting probability $P(\X)$ of reaching the narrow targets $\p\Omega_{hj}$ before escaping from the dendrite. It is defined by
\beq
P(\X) = \text{Prob}\left\{ \tau_{\X}^A < \tau_{\X}^{R_d}  \right\}\,,
\eeq
where $\tau_{\X}^A$ and $\tau_{\X}^{R_d}$ are random times
\begin{align}
\tau_{\X}^A &= \inf \left\{ \tau \, \hbox{such that} \, W(\tau) \in \cup_{j=1}^{n_r}\p\Omega_{hj} | W(0) = \X \right\}\,, \\
\tau_{\X}^{R_d} &= \inf \left\{ \tau \, \hbox{ such that }  \, W(\tau) \in \cup_{j=1}^2\p\Omega_{dj} |W(0) = \X \right\}\,,
\end{align}
where $W(\tau)$ is the underlying Brownian motion. The splitting probability $P(\X)$ is solution of the Laplace's equation \cite{karlin1982diffusion,taflia2007,delgado2015}
\beq\label{eq:prob_pde_dim}
D\Delta P(\X) = 0\,, \quad \X \in \Omega\,,
\eeq
with mixed boundary conditions
\beq\label{eq:prob_bc_dim}
\left\{
\begin{split}
&P(\X) = 1\,, \quad \X \in \cup_{j=1}^{n_r}\p\Omega_{hj}\,, \quad \left[(-1)^jD\frac{\p}{\p X} + \Lambda\right] P(\X) \to 0\,, \quad \X \in \p\Omega_{dj}\,, \\
&D\frac{\p}{\p\n}P(\X) \to 0\,, \quad \X \in \p\Omega\backslash\left\{ \cup_{j=1}^{n_r}\p\Omega_{hj} \cup\p\Omega_{d1}\cup\p\Omega_{d2} \right\}\,,
\end{split}
\right.
\eeq
where $D$ is the diffusion coefficient of the underlying Brownian motion, $\Lambda$ measures the permeability of the boundary caps $\p\Omega_{dj}$, and $\n$ is the unit normal vector to the boundary $\p\Omega$. The Conditional Mean First Passage Time $T(\X)$ is defined by
\beq\label{eq:cond_pde_dim}
T(\X) = E\left\{ \tau_{\X}^A \, | \, \tau_{\X}^A < \tau_{\X}^{R_d}  \,, \X(0)=\X\right\}
\eeq
and satisfies \cite{karlin1982diffusion,taflia2007,delgado2015}
\beq
D\Delta [P(\X)T(\X)] = - P(\X)\,, \quad \X \in \Omega\,,
\eeq
with boundary conditions
\beq\label{eq:cond_bc_dim}
\left\{
\begin{split}
&P(\X)T(\X) = 0\,, \quad \X \in \cup_{j=1}^{n_r}\p\Omega_{hj}\,, \quad \left[(-1)^jD\frac{\p}{\p X} + \Lambda\right] [P(\X)T(\X)] \to 0\,, \quad \X \in \p\Omega_{dj}\,, \quad j=1,2\,, \\
&D\frac{\p}{\p\n}[P(\X)T(\X)] \to 0\,, \quad \X \in \p\Omega\backslash\left\{ \cup_{j=1}^{n_r}\p\Omega_{hj} \cup\p\Omega_{d1}\cup\p\Omega_{d2} \right\}\,,
\end{split}
\right.
\eeq
We will now either assume that the boundary of the dendritic domain is reflective, for which the permeability parameter vanishes $\Lambda = 0$, or we will set $\Lambda = \infty$ thereby yielding perfectly absorbing planar boundaries of the dendritic cylinder.
\subsection{Non-dimensionalization and matching conditions}
We define the dimensionless variables by normalizing with the radius $R$ of the spherical domain $\Omega_h$, so that
\beq
\x = (x,y,z) = \frac{\X}{R}\,, \quad p(\x) = \frac{P(R\x)}{P_0}\,, \quad t(\x) = \frac{D}{R^2}T(R\x)\,, \quad \text{for } \x \in \omega = \frac{\Omega}{R}\,,
\eeq
where $P_0 \equiv 1$ is the characteristic splitting probability, and
\beq
\omega = \omega_h \cup \omega_n \cup \omega_d
\eeq
is the rescaled composite domain with associated dimensionless geometrical parameters
\beq
a = \frac{A}{R}\,, \quad r_n = \frac{R_n}{R}\,, \quad l_n = \frac{L_n}{R}\,, \quad r_d = \frac{R_d}{R}\,, \quad l_d = \frac{L_d}{R}\,.
\eeq
Then by defining the dimensionless membrane permeability as $\lambda = R\Lambda/D$ the mixed boundary-value problems \eqref{eq:prob_pde_dim}-\eqref{eq:cond_bc_dim} become
\beq\label{eq:mixed_bvp1_ndim}
\left\{
\begin{split}
&\Delta p(\x) = 0\,, \quad \x \in \omega\,, \quad p(\x) = 1\,, \quad \x \in \cup_{j=1}^{n_r}\p\omega_{hj}\,, \\
&\left[(-1)^j\frac{\p}{\p x} + \lambda\right] p(\x) = 0\,, \quad \x \in \p\omega_{dj}\,, \quad \frac{\p}{\p\n}p(x) = 0\,, \quad \x \in \p\omega\backslash\left\{ \cup_{j=1}^{n_r}\p\omega_{hj} \cup\p\omega_{d1}\cup\p\omega_{d2} \right\}\,,
\end{split}
\right.
\eeq
for the splitting probability $p(\x)$. Similarly, the normalized conditional time satisfies
\beq\label{eq:mixed_bvp2_ndim}
\left\{
\begin{split}
&\Delta [p(\x)t(\x)] = -p(\x)\,, \quad \x \in \omega\,, \quad p(\x)t(\x) = 0\,, \quad \x \in \cup_{j=1}^{n_r}\p\omega_{hj}\,, \\
&\left[(-1)^j\frac{\p}{\p x} + \lambda\right] [p(\x)t(\x)] = 0\,, \quad \x \in \p\omega_{dj}\,, \quad \frac{\p}{\p\n}[p(\x)t(\x)] = 0\,, \quad \x \in \p\omega\backslash\left\{ \cup_{j=1}^{n_r}\p\omega_{hj} \cup\p\omega_{d1}\cup\p\omega_{d2} \right\}\,,
\end{split}
\right.
\eeq
for the conditional mean first passage time $t(\x)$. We shall use the function
\beq\label{eq:mixed_bvp3_ndim}
v(\x)=t(\x){p(\x)}\,,
\eeq
such that the mixed boundary value problem \eqref{eq:mixed_bvp2_ndim} becomes
\beq
\left\{
\begin{split}
&\Delta v(\x) = -p(\x)\,, \quad \x \in \omega\,, \quad v(\x) = 0\,, \quad \x \in \cup_{j=1}^{n_r}\p\omega_{hj}\,, \\
&\left[(-1)^j\frac{\p}{\p x} + \lambda\right] v(\x) = 0\,, \quad \x \in \p\omega_{dj}\,, \quad \frac{\p}{\p\n}v(\x) = 0\,, \quad \x \in \p\omega\backslash\left\{ \cup_{j=1}^{n_r}\p\omega_{hj} \cup\p\omega_{d1}\cup\p\omega_{d2} \right\}\,. \\
\end{split}
\right.
\eeq
Next we decompose \eqref{eq:mixed_bvp1_ndim} and \eqref{eq:mixed_bvp3_ndim} into three subproblems on each specific domain section $\omega_d$, $\omega_n$ and $\omega_h$. Within the cylindrical dendrite $\omega_d$, we look for a solution of
\beq
\Delta p_d(\x) = 0\,, \quad \Delta v_d(\x) = -p_d(\x)\,, \quad \text{for } \x \in \omega_d\,,
\eeq
satisfying the boundary conditions
\beq
\left\{
\begin{split}
&\frac{\p}{\p \n}p_d(\x) = 0\,, \quad \frac{\p}{\p \n}v_d(\x) = 0\,, \quad \x \in \p\omega_d\backslash\left\{\p\omega_n\cup\p\omega_{d1}\cup\omega_{d2}\right\}\,, \\
&\left[ (-1)^j \frac{\p}{\p x} + \lambda \right]p_d(\x) = 0\,, \quad \left[ (-1)^j \frac{\p}{\p x} + \lambda \right]v_d(\x) = 0\,, \quad \x \in \p\omega_{dj}\,.
\end{split}
\right.
\eeq
At the intersection boundary between the dendrite and the neck, we impose continuity matching conditions:
\beq
p_n(\x) = p_d(\x) \quad \text{and} \quad v_n(\x) = v_d(\x)\,, \quad x \in \p\omega_n\cap\p\omega_d\,,
\eeq
and similarly the flux should be continuous
\beq
\frac{\p}{\p \n}p_d(\x) = \left.\frac{\p }{\p z}p_n(\x)\right|_{z=0} \quad \text{and} \quad \frac{\p}{\p \n}v_d(\x) = \left.\frac{\p}{\p z}v_n(\x)\right|_{z = 0}\,, \quad \x \in \p\omega_n\cap\p\omega_d\,.
\eeq
Here $p_n(\x)$ and $v_n(\x)$ are the splitting probability and within the neck section we have
\beq
\Delta p_n(\x) = 0\,, \quad \Delta v_n(\x) = -p_n(\x)\,, \quad \text{for } \x \in \omega_n\,,
\eeq
along with no-flux boundary conditions on the curved cylindrical boundary,
\beq
\frac{\p}{\p \n}p_n(\x) = 0\,, \quad \frac{\p}{\p \n}v_n(\x) = 0\,, \quad \x \in \p\omega_n\backslash\left\{\p\omega_d\cup\p\omega_h\right\}\,.
\eeq
Additional matching conditions are imposed at the top of the neck, where it intersects with the spherical head,
\beq
\left\{
\begin{split}
p_n(\x) &= p_h(\x) \quad \text{and} \quad v_n(\x) = v_h(\x)\,, \quad x \in \p\omega_n\cap\p\omega_h\,, \\
\frac{\p}{\p \n}p_h(\x) &= \left.-\frac{\p }{\p z}p_n(\x)\right|_{z=l_n} \quad \text{and} \quad \frac{\p}{\p \n}v_h(\x) = \left.-\frac{\p}{\p z}v_n(\x)\right|_{z = l_n}\,, \quad \x \in \p\omega_n\cap\p\omega_h\,.
\end{split}
\right.
\eeq
where $p_h(\x)$ and $v_h(\x)$ are the splitting probability and the function $ v_h$ restricted to the head section $\omega_h$ satisfies
\beq
\Delta p_h(\x) = 0\,, \quad \Delta v_h(\x) = -p_h(\x)\,, \quad \text{for } \x \in \omega_h\,,
\eeq
subject to mixed Neumann-Dirichlet boundary conditions
\beq
\left\{
\begin{split}
&\frac{\p}{\p \n}p_h(\x) = 0\,, \quad \frac{\p}{\p \n}v_h(\x) = 0\,, \quad \x \in \p\omega_h\backslash\left\{\cup_{j=1}^{n_r}\p\omega_{hj}\cup\p\omega_n\right\}\,, \\
&p_h(\x) = 1\,, \quad v_h(\x) = 0\,, \quad \x \in \cup_{j=1}^{n_r} \p \omega_{hj}\,.
\end{split}
\right.
\eeq
\subsection{Fully reflective dendrite boundaries}\label{sec:no_leak}
When the membrane permeability ratio $\lambda \equiv R\Lambda/D = 0$ vanishes, diffusing particles cannot escape from the dendrite and thus the splitting probability is $p(\x) \equiv 1$. We first define
\beq
\x_d = (0,0,0)\,, \quad \x_h = (0,0,l_n)\,,
\eeq
as the centers of the circular disks $\p\omega_n\cap\p\omega_d$ and $\p\omega_n\cap\p\omega_h$ connecting the narrow passage of radius $r_n$ between the dendrite and the spherical head. Next, the solution $v_n(z)$ in the thin cylindrical neck is radially symmetric and satisfies the following boundary value problem
\beq
\begin{split}
&\frac{d^2}{dz^2}v_n(z) = -1\,, \quad 0 \leq z \leq l_n\,, \quad v_n(0) = v_d(\x_d)\,, \quad v_n(l_n) = v_h(\x_h)\,, \\
&\left.\frac{d}{dz}v_n\right|_{z=0} = \frac{\p}{\p\n}v_d(\x)\,, \quad \text{for } \x \in \p\omega_n\cap\p\omega_d\,, \quad \left.\frac{d}{dz}v_n\right|_{z=l_n} = - \frac{\p}{\p\n}v_h(\x) \,, \quad \text{for } x \in \p\omega_n\cap\p\omega_h\,,
\end{split}
\eeq
where $z$ corresponds to the distance from the dendrite. The solution is readily found to be
\beq
v_n(z) = v_d(\x_d) + \left(\frac{1}{l_n}\left(v_h(\x_h) - v_d(\x_d)\right) + \frac{l_n}{2}\right)z - \frac{z^2}{2}\,.
\eeq
Next, by applying the divergence theorem we show in Appendix \ref{sec:dendrite} that the exit flux from the dendrite must satisfy
\beq
\frac{\p}{\p\n}v_d(\x) = - \frac{|\omega_d|}{\pi r_n^2}\,, \quad \x \in \p\omega_n\cap\omega_d
\eeq
and thus upon using the matching condition, we get
\beq\label{eq:noleak_eq1}
\left.\frac{d}{dz}v_n\right|_{z=0} = \frac{1}{l_n}\left(v_h(\x_h) - v_d(\x_d)\right) + \frac{l_n}{2} = - \frac{|\omega_d|}{\pi r_n^2}.
\eeq
At the intersection $\p\omega_n\cap\p\omega_h$, we obtain the asymptotic behavior result (Appendix \ref{sec:head}) for small $r_n$:
\beq
v_h(\x_h) = \frac{1}{4an_r}\left(|\omega_h| - \pi r_n^2 \left.\frac{d}{dz}v_n\right|_{z=l_n}\right)\left(1 - \frac{a}{\pi}\log(a) + O(a)\right) - \left.\frac{d}{dz}v_h\right|_{z=l_n}\left(r_n - \frac{r_n^2}{4}\log(r_n) + O(r_n^2)\right)\,,
\eeq
which becomes upon using the matching condition
\beq\label{eq:noleak_eq2}
\begin{split}
v_h(\x_h) &= \left(\frac{l_nr_n}{2} - \frac{r_n}{l_n}\left(v_h(\x_h) - v_d(\x_d)\right)\right)\left(\frac{\pi r_n}{4an_r}\left(1 - \frac{a}{\pi}\log(a) + O(1)\right) + 1 - \frac{r_n}{4}\log(r_n) + O(r_n)\right) \\
&+ \frac{|\omega_h|}{4an_r}\left(1 - \frac{a}{\pi}\log(a) + O(a)\right).
\end{split}
\eeq
Then, upon solving \eqref{eq:noleak_eq1} and \eqref{eq:noleak_eq2} for $v_d(\x_d)$ and $v_h(\x_h)$, we obtain
\begin{align*}
v_d(\x_d) &= \frac{|\omega|}{4an_r}\left(1 - \frac{a}{\pi}\log(a) + O(a)\right) + l_nr_n\left(\frac{|\omega_d|}{|\omega_n|} + 1\right)\left(1 - \frac{r_n}{4}\log(r_n) + O(r_n)\right) + l_n^2\left(\frac{1}{2} + \frac{|\omega_d|}{|\omega_n|}\right)\,, \\
v_h(\x_h) &= \frac{|\omega|}{4an_r}\left(1 - \frac{a}{\pi}\log(a) + O(a)\right) + l_nr_n\left(\frac{|\omega_d|}{|\omega_n|} + 1\right)\left(1 - \frac{r_n}{4}\log(r_n) + O(r_n)\right)\,,
\end{align*}
where $|\omega|$ is the volume of the composite domain
\beq
|\omega| = |\omega_h| + |\omega_n| + |\omega_d| = \frac{4\pi}{3} + \pi r_n^2l_n + \pi r_d^2l_d\,,
\eeq
and thus within the neck $v_n(z)$, we have
\beq\label{eq:v_n_z}
\begin{split}
v_n(z) &= \frac{|\omega|}{4an_r}\left(1 - \frac{a}{\pi}\log(a) + O(a)\right) + l_nr_n\left(\frac{|\omega_d|}{|\omega_n|} + 1\right)\left(1 - \frac{r_n}{4}\log(r_n) + O(r_n)\right) \\
&+l_n^2\left(\frac{1}{2} + \frac{|\omega_d|}{|\omega_n|}\right) - z^2\left(\frac{|\omega_d|}{|\omega_n|} \frac{l_n}{z} + \frac{1}{2}\right)\,.
\end{split}
\eeq
Since the splitting probability satisfies $p_n(z) = 1$ we simply need to multiply \eqref{eq:v_n_z} by the factor $R^2/D$ to obtain the formula for the MFPT $T_n(Z)$ with all the dimensional parameters,
\beq\label{eq:Lambda_0}
\begin{split}
T_n(Z) &= \frac{|\Omega|}{4Dn_rA}\left(1 - \frac{A}{\pi R}\log\left(\frac{A}{R}\right) + O\left(\frac{A}{R}\right)\right) + \frac{L_dR_d^2}{DR_n}\left(1 + \frac{L_nR_n^2}{L_dR_d^2}\right)\left(1 - \frac{R_n}{4R}\log\left(\frac{R_n}{R}\right) + O\left(\frac{R_n}{R}\right)\right) \\
&+ \frac{L_nL_dR_d^2}{DR_n^2}\left(1 + \frac{L_nR_n^2}{2L_dR_d^2}\right) - \frac{ZL_dR_d^2}{DR_n^2}\left(1 + \frac{ZR_n^2}{2L_dR_d^2}\right) \,, \quad \text{for } 0 \leq Z \leq L_n\,,
\end{split}
\eeq
where $|\Omega| = \frac{4\pi R^3}{3} + \pi R_n^2L_n + \pi R_d^2L_d$ is the volume of the entire composite domain.
\subsection{Computing the splitting probability when particles in the dendrite can be lost}
We now analyze the case where Brownian particles are absorbed by the circular planar boundaries of the dendrite, which is obtained by taking the surface permeability to be $\lambda = \infty$. We study the fraction of these particles that reaches a narrow target (the splitting probability). Given that they have reached the targets, we compute the conditional mean first passage time.\\
We first seek an asymptotic approximation for the splitting probability $p(\x)$. In the thin cylindrical neck $\Omega_n$, the splitting probability $p_n(z)$ satisfies the boundary value problem
\beq
\left\{
\begin{split}
&\frac{d^2}{dz^2}p_n(z) = 0\,, \quad 0 \leq z \leq l_n\,, \quad p_n(0) = p_d(\x_d)\,, \quad p_n(l_n) = p_h(\x_h)\,, \\
&\left.\frac{d}{dz}p_n\right|_{z=0} = \frac{\p}{\p\n}p_d(\x)\,, \quad \x \in \p\omega_n\cap\p\omega_d \quad \left.\frac{d}{dz}p_n\right|_{z=l_n} = -\frac{\p}{\p\n}p_h(\x)\,, \quad \x \in \p\omega_n\cap\p\omega_h\,,
\end{split}
\right.
\eeq
which readily solves as
\beq
p_n(z) = \left(p_h(\x_h) - p_d(\x_d)\right)\frac{z}{l_n} + p_d(\x_d)\,, \quad 0 \leq z \leq l_n\,.
\eeq
Here $p_d(\x_d)$ and $p_h(\x_h)$ are the splitting probabilities at the bottom and top sections of the neck, whose expressions are obtained by an asymptotic expansion (Appendix \ref{sec:head} and \ref{sec:dendrite}). The results are
\beq\label{eq:asym1}
p_d(\x_d) = \left.\frac{d}{dz}p_n\right|_{z=0} r_n = \left(p_h(\x_h) - p_d(\x_d)\right)\frac{r_n}{l_n}\,,
\eeq
and
\beq\label{eq:asym2}
p_h(\x_h) = 1 - \left.\frac{d}{dz}p_n\right|_{z=l_n}\left( \frac{\pi r_n^2}{4an_r} + r_n\right) = 1 - \frac{r_n}{l_n}\left(p_h(\x_h) - p_d(\x_d)\right)\left( \frac{\pi r_n}{4an_r} + 1\right)\,,
\eeq
where we only kept the leading order terms in $a \ll 1$ and $r_n \ll 1$. Upon solving equations \eqref{eq:asym1} and \eqref{eq:asym2} we find
\beq
p_d(\x_d) = \frac{1}{\mu}\,, \quad p_h(\x_h) = \frac{1 + \frac{l_n}{r_n}}{\mu}\,, \quad \text{with} \quad \mu = 2 + \frac{l_n}{r_n} + \frac{\pi r_n}{4an_r}\,.
\eeq
and thus we get within the neck
\beq\label{eq:p_n_z}
p_n(z) = \frac{1 + \frac{z}{r_n}}{\mu}\,, \quad 0 \leq z \leq l_n\,.
\eeq
The probability of reaching a single specific target $\p\omega_{hj}$ is recovered after dividing by the total number $n_r$. Finally, the dimensional formula associated to \eqref{eq:p_n_z} is given by
\beq
P_n(Z) = \frac{1 + \frac{Z}{R_n}}{2 + \frac{L_n}{R_n} + \frac{\pi R_n}{4An_r}} + O\left(\left(\frac{R_n}{R}\right)^2\log\left(\frac{A}{R}\right)\right)\,, \quad \text{for } 0 \leq Z \leq L_n\,.
\eeq
\subsection{Conditional MFPTs for particles absorbed by the dendrite}
We now solve for the conditional mean first passage time in the thin cylindrical neck limit: the  function $v_n(z)$ satisfies the boundary value problem
\beq\label{eq:cond_neck}
\left\{
\begin{split}
&\frac{d^2}{dz^2}v_n(z) = -p_n(z) = - \frac{1 + \frac{z}{r_n}}{\mu}\,, \quad 0 \leq z \leq l_n\,, \quad v_n(0) = v_d(\x_d)\,, \quad v_n(l_n) = v_h(\x_h)\,, \\
&\left.\frac{d}{dz}v_n\right|_{z=0} = \frac{\p}{\p\n}v_d(\x)\,, \quad \x \in \p\omega_n\cap\p\omega_d\,, \quad \left.\frac{d}{dz}v_n\right|_{z=l_n} = -\frac{\p}{\p\n}v_h(\x)\,, \quad \x \in \p\omega_n\cap\p\omega_h \,,
\end{split}
\right.
\eeq
where $v_d(\x_d)$ and $v_h(\x_h)$ are solutions within the dendrite and head compartments respectively. After integrating with respect to the distance $Z$ from the dendrite, we get
\beq
v_n(z) = v_d(\x_d) + \left(\frac{1}{l_n}\left(v_h(\x_h) - v_d(\x_d)\right) + \frac{\frac{l_n}{2} + \frac{l_n^2}{6r_n}}{\mu} \right)z - \left(\frac{\frac{z^2}{2} + \frac{z^3}{6r_n}}{\mu} \right)\,, \quad 0 \leq z \leq l_n.
\eeq
Keeping the leading order terms in $a$ and $r_n$ from equations \eqref{eq:vhxh} and \eqref{eq:asymptotic_result_dendrite}, we get
\beq\label{eq:cond_asym_1}
v_d(\x_d) = \left.\frac{d}{dz}v_n\right|_{z=0} r_n = \left(v_h(\x_h) - v_d(\x_d)\right)\frac{r_n}{l_n} + \frac{\frac{l_nr_n}{2} + \frac{l_n^2}{6}}{\mu}
\eeq
as well as
\begin{align}\label{eq:cond_asym_2}
v_h(\x_h) &= \frac{1}{4an_r}\left( |\omega_h|\overline{p_h} - \left.\frac{d}{dz}v_n\right|_{z=l_n}\pi r_n^2 \right) - \left.\frac{d}{dz}v_n\right|_{z=l_n}r_n \nonumber \\
&= \frac{|\omega_h|\overline{p_h}}{4an_r} + \left(\frac{\frac{l_nr_n}{2} + \frac{l_n^2}{3}}{\mu} - \frac{r_n}{l_n}\left(v_h(\x_h)-v_d(\x_d)\right)\right) \left(\frac{\pi r_n}{4an_r} + 1\right) \,.
\end{align}
Here $\overline{p_h}$ is the average splitting probability over the head domain $\omega_h$, whose leading order approximation is
\beq
\overline{p_h} = 1 - \frac{\pi r_n^2}{4an_r}\left.\frac{d}{dz}p_n\right|_{z=l_n} = 1 - \frac{\pi r_n}{4an_r \mu} = \frac{2 + \frac{l_n}{r_n}}{\mu}\,,
\eeq
as obtained from formula \eqref{eq:phbar} of Appendix \ref{sec:head}. Then, upon solving \eqref{eq:cond_asym_1} and \eqref{eq:cond_asym_2} for $v_d(\x_d)$ and $v_h(\x_h)$, we obtain
\beq
v_d(\x_d) = \frac{1}{\mu}\left( \frac{|\omega_h|}{4an_r}\left(\frac{2 + \frac{l_n}{r_n}}{\mu} + \frac{\pi r_nl_n^2}{2|\omega_h|\mu} + \frac{\pi r_n^2 l_n}{|\omega_h|\mu}\right) + \frac{l_n^3}{6r_n\mu} + \frac{l_n^2}{\mu} + \frac{r_nl_n}{\mu}\right)\,,
\eeq
as well as
\beq
v_h(\x_h) = \frac{1 + \frac{l_n}{r_n}}{\mu}\left( \frac{|\omega_h|}{4an_r}\left(\frac{2 + \frac{l_n}{r_n}}{\mu} + \frac{\pi r_nl_n^2}{2|\omega_h|\mu} + \frac{\pi r_n^2 l_n}{|\omega_h|\mu}\right) + \frac{l_n^3}{6r_n\mu} + \frac{l_n^2}{\mu} + \frac{r_nl_n}{\mu}\right) - \frac{\frac{l_n^2}{2} + \frac{l_n^3}{6r_n}}{\mu}\,.
\eeq
For $0 \leq z \leq l_n$, this therefore yields
\beq
v_n(z) = \frac{1 + \frac{z}{r_n}}{\mu}\left( \frac{|\omega_h|}{4an_r}\left(\frac{2 + \frac{l_n}{r_n}}{\mu} + \frac{3r_nl_n^2}{8\mu} + \frac{3r_n^2 l_n}{4\mu}\right) + \frac{l_n^3}{6r_n\mu} + \frac{l_n^2}{\mu} + \frac{r_nl_n}{\mu}\right) - \left(\frac{\frac{z^2}{2} + \frac{z^3}{6r_n}}{\mu} \right)\,,
\eeq
where the volume of the head is $|\omega_h| = 4\pi/3$. To recover the conditional mean first passage times for an arbitrary starting point within the neck, we divide by the splitting probability to obtain
\beq
t_n(z) = \frac{v_n(z)}{p_n(z)} = \left( \frac{|\omega_h|}{4an_r}\left(\frac{2 + \frac{l_n}{r_n}}{\mu} + \frac{3r_nl_n^2}{8\mu} + \frac{3r_n^2 l_n}{4\mu}\right) + \frac{l_n^3}{6r_n\mu} + \frac{l_n^2}{\mu} + \frac{r_nl_n}{\mu}\right) - \frac{z^2}{6}\left(1 + \frac{2}{1 + \frac{z}{r_n}} \right)\,,
\eeq
which reduces to
\beq\label{eq:t_n_z}
t_n(z) = \frac{1}{2 + \frac{l_n}{r_n} + \frac{\pi r_n}{4an_r}}\left( \frac{|\omega_h|}{4an_r}\left(2 + \frac{l_n}{r_n} + \frac{3r_nl_n^2}{8} + \frac{3r_n^2 l_n}{4}\right) + \frac{l_n^3}{6r_n} + l_n^2 + r_nl_n\right) - \frac{z^2}{6}\left(1 + \frac{2}{1 + \frac{z}{r_n}} \right)\,.
\eeq
This yields for a particle starting at the intersection of the neck with the dendrite
\beq
t_d(\x_d) = \frac{1}{2 + \frac{l_n}{r_n} + \frac{\pi r_n}{4an_r}}\left( \frac{|\omega_h|}{4an_r}\left(2 + \frac{l_n}{r_n} + \frac{3r_nl_n^2}{8} + \frac{3r_n^2 l_n}{4}\right) + \frac{l_n^3}{6r_n} + l_n^2 + r_nl_n\right)\,,
\eeq
and for those starting at the upper neck cross section, we have
\beq\label{eq:thxh_ndim}
t_h(\x_h) = \frac{1}{2 + \frac{l_n}{r_n} + \frac{\pi r_n}{4an_r}}\left( \frac{|\omega_h|}{4an_r}\left(2 + \frac{l_n}{r_n} + \frac{3r_nl_n^2}{8} + \frac{3r_n^2 l_n}{4}\right) + \frac{l_n^3}{6r_n} + l_n^2 + r_nl_n\right) - \frac{l_n^2}{6}\left(1 + \frac{2}{1 + \frac{l_n}{r_n}} \right)\,.
\eeq
Finally we set $T_n(Z) = R^2/Dt_n(Z/R)$ and obtain the conditional MFPT formula equivalent to \eqref{eq:t_n_z} with dimensional units:
\beq
\begin{split}
T_n(Z) &= \frac{1}{2 + \frac{L_n}{R_n} + \frac{\pi R_n}{4An_r}}\left(\frac{\pi R^3}{3DAn_r}\left(2 + \frac{L_n}{R_n} + \frac{3R_nL_n^2}{8R^3} + \frac{3R_n^2L_n}{4R^3} \right) + \frac{L_n^2}{D}\left(1 + \frac{L_n}{6R_n} + \frac{R_n}{L_n}\right)\right) \\
&- \frac{Z^2}{6D}\left(1 + \frac{2}{1 + \frac{Z}{R_n}} \right) + O\left(\log\left(\frac{A}{R}\right)\right)\,, \quad \text{for } 0 \leq Z \leq L_n\,.
\end{split}
\eeq
\subsection{Asymptotic formulas versus numerical simulations}
We now compare the asymptotic formulas for the splitting probability and the conditional MFPT against numerical solutions of the mixed boundary value problems \eqref{eq:mixed_bvp1_ndim} and \eqref{eq:mixed_bvp2_ndim} performed with COMSOL \cite{comsol}. Additional stochastic simulations are also performed, but only for the case where Brownian particles are absorbed upon hitting the lateral boundaries of the cylindrical dendrite. All results are shown with dimensional units and the parameter values are summarized in Table \ref{tableS2}. \\
We first consider in Fig.~\ref{fig:Lambda_0} the case where no particles are lost from the dendrite, i.e.\ $\Lambda = 0$ and the splitting probability is $P(\X) = 1$. Then we set $\Lambda = \infty$ and analyze how the narrow exits radius, as well as the length and the radius of the cylindrical neck, affect the splitting probability and the conditional MFPT (Fig.~\ref{fig:Lambda_inf_fig1}-\ref{fig:Lambda_inf_fig3}). Finally for all the numerical experiments either $n_r = 1,\,5$ or $50$ targets are added to the spherical boundary $\p\Omega_h$.
\begin{table}[http!]
\caption{\label{tableS2} \textbf{List of parameter values}}
\vspace{1em}
\centering
\begin{tabular}{lcrcr}
\hline			
\textbf{Parameter}  &   \textbf{Symbol}    &   \textbf{Value}   &   \textbf{Non-dim.\ parameter}    &   \textbf{Non-dim.\ value}\\
\hline			
Spine head radius    		& $R$			& 0.5 $\mu$m     & $R/R$   & $1$\\
Spine neck radius  	    	& $R_n$	 		& 0.15 $\mu$m   & $r_n = R_n/R$ & $0.3$ \\
Spine neck length           & $L_n$			& 0.5 $\mu$m  & $l_n = L_n/R$ & $1$ \\
Dendrite radius             & $R_d$	 		& 1 $\mu$m   & $r_d = R_d/R$ & $2$ \\
Dendrite length     		& $L_d$			& 5 $\mu$m  & $l_d = L_d/R$ & $10$ \\
Circular target radius  	& $A$	 		& 0.01 $\mu$m & $a = A/R$ & $0.02$ \\
Number of targets & $n_r$			& $50$ & $-$ & $-$ \\
Diffusion coefficient 		& $D$ 			& 350 $\mu {\rm m}^2{\rm s}^{-1}$ & $-$ & $-$ \\
\hline
\end{tabular}
\end{table}
\subsection{Case $\Lambda = 0$}
\begin{figure}[http!]
\centering
\includegraphics[width=\textwidth]{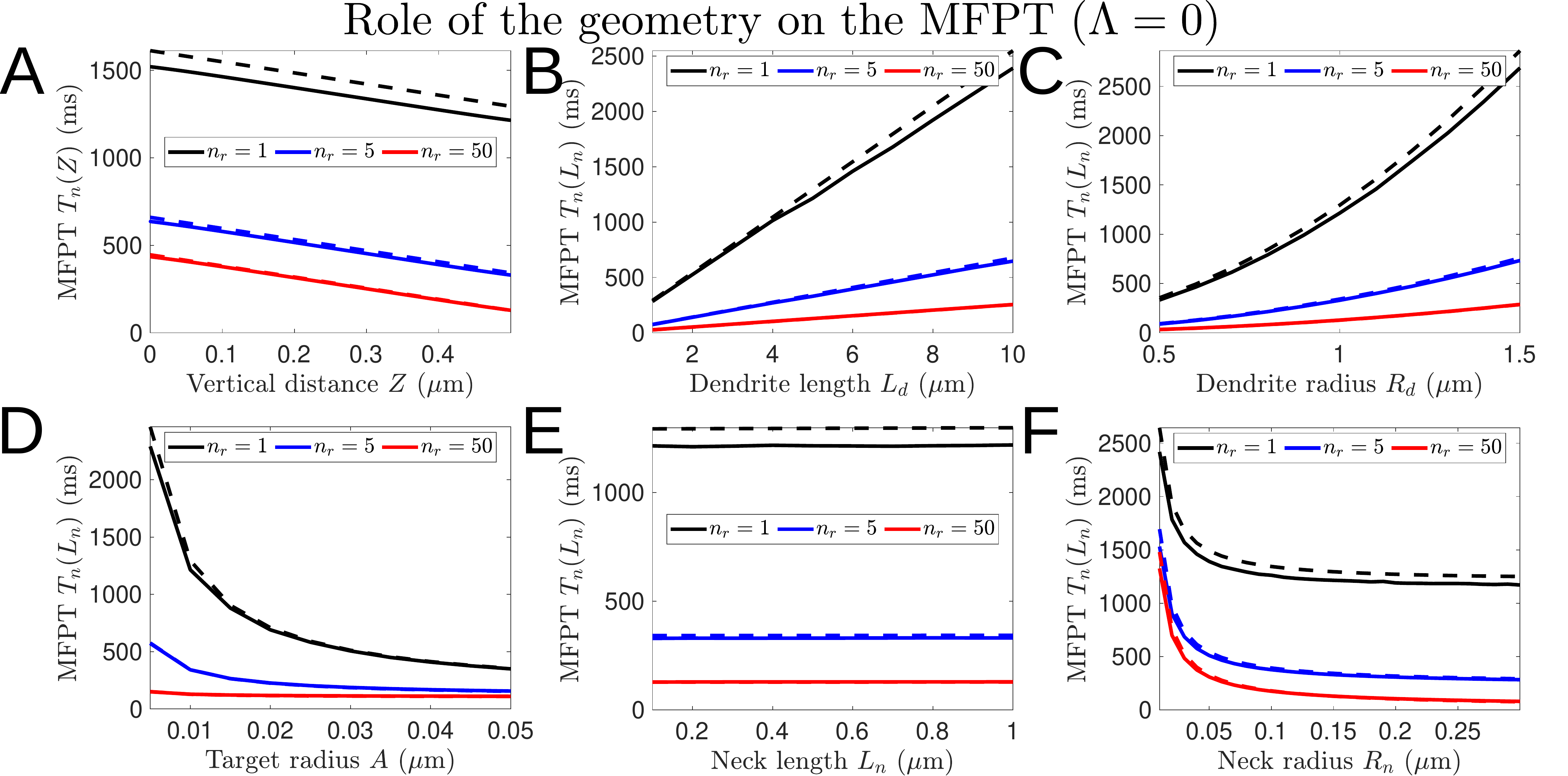}
\caption{\label{fig:Lambda_0} \textbf{Effects of the geometry on the MFPT, case $\Lambda = 0$.} (\textbf{A}): MFPT $T_n(Z)$ as a function of the distance $Z$ from the dendrite. (\textbf{B})-(\textbf{C}): We set $Z=L_n$ and vary the length $L_d$ and the radius $R_d$ of the cylindrical dendrite. (\textbf{D}): The MFPT $T_n(L_n)$ is instead shown as a function of the target radius $A$. (\textbf{E})-(\textbf{F}): Parameter sweeps of the narrow cylindrical neck of length $L_n$ and radius $R_n$ are performed. The dashed and full curves correspond to the asymptotic and COMSOL numerical solutions.}
\end{figure}

When $\Lambda = 0$ the splitting probability is $P(\X) = 1$ and Brownian particles can only escape from the dendrite through the narrow cylindrical neck. We show that adding more targets on the head boundary $\p\Omega_h$ reduces the average search times by an equivalent factor: when they are well-spaced, twice as many targets halve the MFPT. Also the average search time decreases linearly as the starting point $Z$ approaches the spherical head (Fig.~\ref{fig:Lambda_0}\textbf{A}). We then set the initial position at $Z = L_n$ and perform a sweep of the different parameters characterizing the geometry of the dendritic spine (Fig.~\ref{fig:Lambda_0}\textbf{B}-\textbf{F}). Our asymptotic analysis predicts that increasing the volume of the dendrite yields longer average escape times (Fig.~\ref{fig:Lambda_0}\textbf{B}-\textbf{C}). We recover the usual $1/A$ reciprocity relation \cite{cheviakov2010} as the targets radius $A$ increases, as shown in Fig.~\ref{fig:Lambda_0}\textbf{D}.\\
Finally, we report in Fig.~\ref{fig:Lambda_0}\textbf{E}-\textbf{F} that the MFPT $T_n(L_n)$ is not significantly affected by neck length variations, while we obtain the $1/R_n^2$ dependency as the radius of the narrow passage is increased, a relation also obtained for calcium ions that are reaching the bottom of the neck \cite{biess2007diffusion,holcman2011diffusion}.
In each plot, we find a good agreement between the asymptotic solution (dashed curve) and the numerical solution (full curve).
\subsection{Case $\Lambda = \infty$}
\begin{figure}[http!]
\centering
\includegraphics[width=\textwidth]{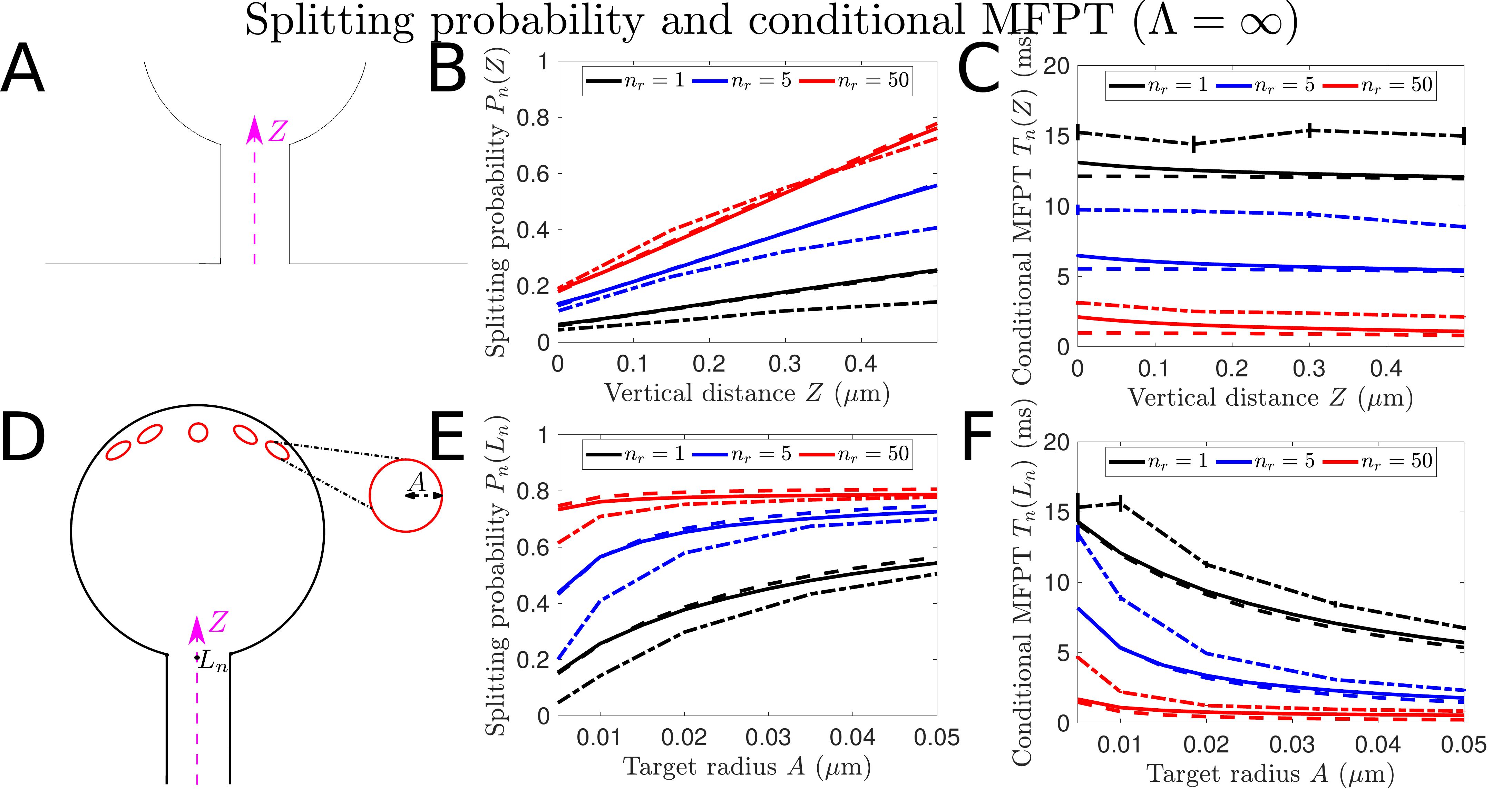}
\caption{\label{fig:Lambda_inf_fig1} The top panels give the splitting probability $P_n(Z)$ (\textbf{B}) and the conditional MFPT $T_n(Z)$ (\textbf{C}) within the narrow cylindrical neck, as a function of the distance $Z$ from the dendrite. In panels (\textbf{D})-(\textbf{F}), we set $Z=L_n$ and show the same quantity as a function of the narrow target radius $A$. In each panel we show the asymptotic solution (dashed-curved), the COMSOL numerical solution (full curve) and the stochastic simulation results (dot-dashed curve, with the vertical bars indicating the standard deviation) for three target numbers: $n_r = 1$ (black), $n_r = 5$ (blue) and $n_r = 50$ (red).}
\end{figure}

\begin{figure}[http!]
\centering
\includegraphics[width=\textwidth]{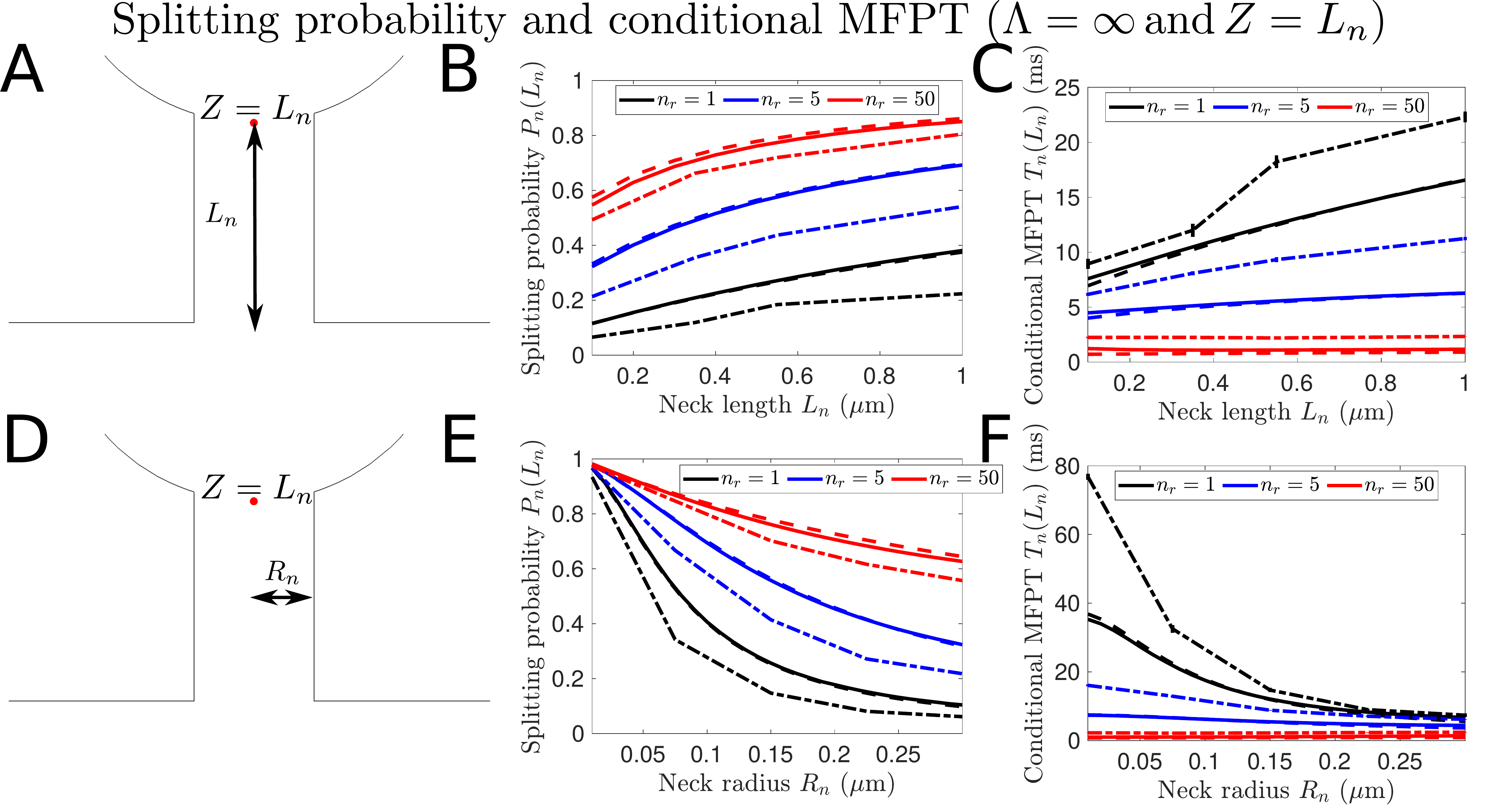}
\caption{\label{fig:Lambda_inf_fig2} In the top panels we set $Z=L_n$ and give the splitting probability $P_n(L_n)$ (\textbf{B}) and the conditional MFPT $T_n(L_n)$ (\textbf{C}) as a function of the neck length $L_n$. In panels (\textbf{D})-(\textbf{F}) we show the same quantity as a function of the neck radius $R_n$. In each panel we show the asymptotic solution (dashed-curved), the COMSOL numerical solution (full curve) and the stochastic simulation results (dot-dashed curve, with the vertical bars indicating the standard deviation) for three target numbers: $n_r = 1$ (black), $n_r = 5$ (blue) and $n_r = 50$ (red).}
\end{figure}

\begin{figure}[http!]
\centering
\includegraphics[width=\textwidth]{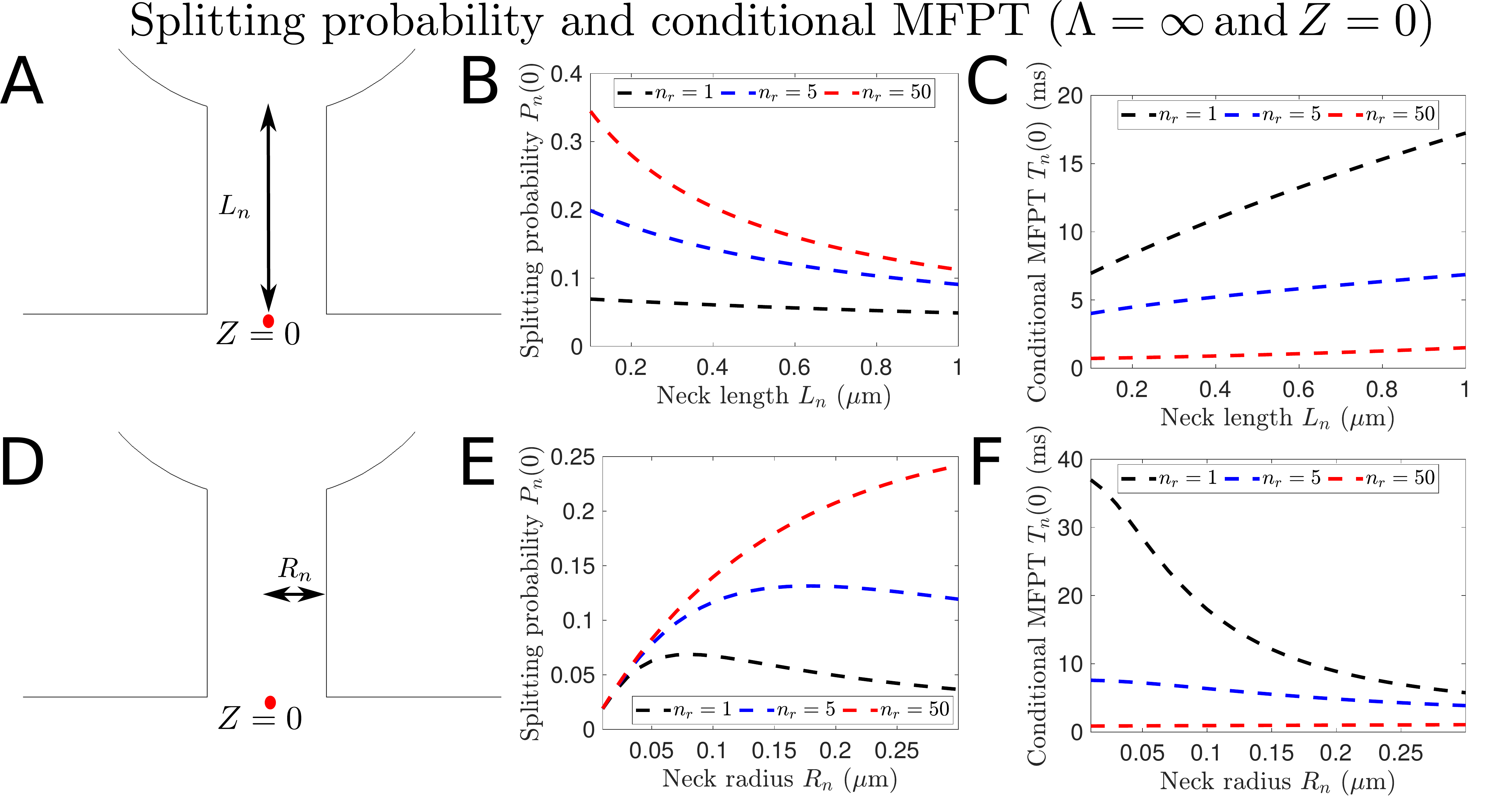}
\caption{\label{fig:Lambda_inf_fig3} In the top panels we set $Z=0$ and give the splitting probability $P_n(0)$ (\textbf{B}) and the conditional MFPT $T_n(0)$ (\textbf{C}) as a function of the neck length $L_n$. In panels (\textbf{D})-(\textbf{F}) we show the same quantity as a function of the neck radius $R_n$. In each panel we only show the asymptotic solution (dashed-curved) for three target numbers: $n_r = 1$ (black), $n_r = 5$ (blue) and $n_r = 50$ (red).}
\end{figure}

We now turn to the case where Brownian particles are absorbed upon hitting the lateral boundaries of the cylindrical dendrite $\p\Omega_{d1}$ and $\p\Omega_{d2}$. First, because we conditioned on the particles reaching any targets on the boundary of the head, the MFPT decays drastically from a few hundreds to a few tens of milliseconds when comparing both $\Lambda = 0$ and $\Lambda = \infty$ cases. We find that the splitting probability increases linearly with the distance $Z$ from the dendrite, as shown in Fig.~\ref{fig:Lambda_inf_fig1}\textbf{B}. However this is little variations of the conditional MFPT (Fig.~\ref{fig:Lambda_inf_fig1}\textbf{C}).\\
The splitting probability increases with the target radius $A$, while the conditional MFPT behaves as $O(1/A)$ as expected from the asymptotic formula \eqref{eq:thxh_ndim} (Fig.~\ref{fig:Lambda_inf_fig1}\textbf{E})-\textbf{F}).\\
Finally, we analyze how the geometry of the narrow cylindrical neck affects the splitting probability and the conditional MFPT (Fig.~\ref{fig:Lambda_inf_fig2}). Upon setting $Z=L_n$ as the initial position, we find that the splitting probability increases with the neck length $L_n$, while it decreases with the neck radius $R_n$. Similar behaviors are obtained for the conditional MFPT $T_n(L_n)$ in Fig.~\ref{fig:Lambda_inf_fig2}\textbf{C} and \textbf{F}, although we remark that for large number of targets varying the neck dimensions does not influence the mean binding times significantly. \\
Interestingly modifying the starting point from the top to the base of the neck only provides qualitatively different behavior for the splitting probability, as shown in Fig.~\ref{fig:Lambda_inf_fig3}, where we remark that increasing the length of the neck yields a smaller probability of reaching any targets. For small target numbers, we obtain that the splitting probability $P_n(0)$ behaves non-monotonically as the radius of the neck increases (Fig.~\ref{fig:Lambda_inf_fig3}\textbf{E}): there is an intermediate $R_n$ value on the range $[0.01,\, 0.3]\, \mu{\rm m}$ for which the splitting probability is maximized. For a large number of targets this maximum is achieved for physiologically non-realistic neck radius values. \\
Finally, we remark that the stochastic simulation results agree qualitatively with the asymptotic solutions (Fig.~\ref{fig:Lambda_inf_fig1} and \ref{fig:Lambda_inf_fig2}) as well as the numerical solutions of the mixed boundary-value problem obtained with COMSOL. The discrepancy observed tends to shrink for larger target numbers, and also as the target and neck radii increase.
\section{Discussion and concluding remarks}\label{sec:discussion}
\begin{figure}[http!]
\centering
\includegraphics[width=0.66\textwidth]{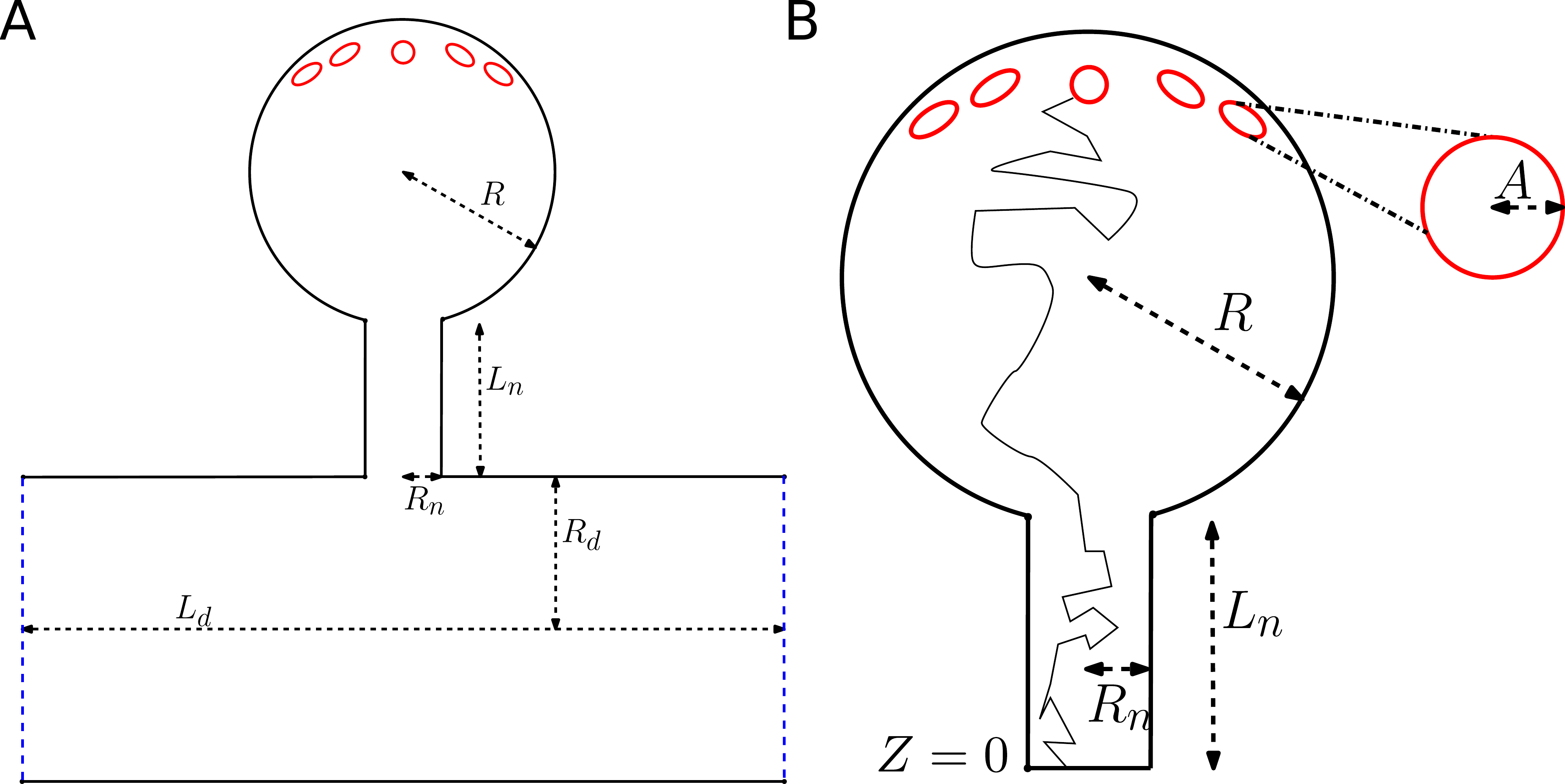}
\caption{\label{fig:mito_new_dendrite} \textbf{Schematic diagram of the dendritic spine.} \textbf{A}: Full geometry with the ball of radius $R$ with the two cylindrical compartments. \textbf{B}: Simplified geometry where the large bottom cylindrical compartment is neglected. The Brownian path in black indicates a successful trajectory reaching a target starting from the neck base.}
\end{figure}

In this manuscript, we modeled diffusion in heterogeneous network with large spherical nodes and thin cylindrical tubules. We derived asymptotic formulas for the mean time a Brownian particle takes to escape from various 3-D composite domains, i.e.\ with both large and short scales, with reflecting boundaries everywhere except for a collection of narrow absorbing windows.  We formulated the mixed Neumann-Dirichlet boundary value problem for the Mean First Passage Time (MFPT) $T(\X)$ on a composite network domain with $M$ by $N$ identical ball compartments connected by narrow cylinders (section \S \ref{sec:net}). \\
Using asymptotic analysis we derived a sparse linear system of equations for the MFPT averaged over each compartment, that can be solved numerically. In the absence of absorbing targets, the linear system is singular and consists of a tridiagonal matrix analogous to the 2-D discrete Laplacian with reflecting boundary conditions. For a single target, there is an additional rank-one perturbation matrix that is linearly proportional to the target radius and narrow passage length, and which scales as the reciprocal of the square of the radius of the narrow passage. Then we considered a composite domain consisting of a sphere with multiple well-spaced absorbing targets (section \S \ref{sec:spine} ), that is connected to a large cylindrical compartment by a narrow cylindrical neck (Fig.~\ref{fig:mito_new_dendrite}\textbf{A}). For Brownian particles starting at the bottom of the neck, we obtain the formula
\beq\label{eq:mfpt1}
\begin{split}
T_n(0) =& \frac{\left(\frac{\pi R^3}{3} + \frac{\pi R_n^2L_n}{4} + \frac{\pi R_d^2L_d}{4}\right)}{Dn_rA}\left(1 - \frac{A}{\pi R}\log\left(\frac{A}{R}\right)\right) + \frac{L_dR_d^2}{DR_n}\left(1 + \frac{L_nR_n^2}{L_dR_d^2}\right)\left(1 - \frac{R_n}{4R}\log\left(\frac{R_n}{R}\right)\right) \\
&+ \frac{L_nL_dR_d^2}{DR_n^2}\left(1 + \frac{L_nR_n^2}{2L_dR_d^2}\right).
\end{split}
\eeq
Thus the MFPT scales with the total volume of the composite domain, as the reciprocal of the targets size, and also as the reciprocal of the square of the radius of the narrow passage.
Using parameter values gathered in Table \ref{tableS2}, we obtain  a numerical value $T_n(0) \approx 450 \, {\rm ms}$ for the diffusion timescale from the dendrite to a small target located in the spine head.\\
When the larger cylindrical compartment (dendrite) is neglected (Fig.~\ref{fig:mito_new_dendrite}\textbf{B}), or simply no passage is possible (imposed by a mitochondria in the context of a neuron), which is equivalent to imposing reflecting boundary conditions at the bottom section of the narrow neck, then the formula \eqref{eq:mfpt1} reduces to
\beq\label{eq:mfpt2}
T_n(0) = \frac{\left(\frac{\pi R^3}{3} + \frac{\pi R_n^2L_n}{4}\right)}{Dn_rA}\left(1 - \frac{A}{\pi R}\log\left(\frac{A}{R}\right)\right) + \frac{L_nR_n}{D}\left(1 - \frac{R_n}{4R}\log\left(\frac{R_n}{R}\right)\right) + \frac{L_n^2}{2D}\,,
\eeq
which leads to the numerical approximation $T_n(0)\approx1.41\,{\rm ms}$, which is nearly 300 times faster than for the full geometry case. Interestingly there is no singular term in $1/R_n^2$ compared to equation \ref{eq:mfpt1}. \\
In case of partial absorption due to the presence of an organelle, we thus expect a time that can be modulated between few to hundreds of milliseconds. This time scale is relevant for quantifying the diffusion of ATP molecules generated by mitochondria and that needs to translocate into dendritic spine to maintain energy integrity. \\
We also analyzed the case where Brownian particles can be absorbed upon hitting the two opposite flat boundaries of the large cylindrical compartment. We derived a formula for the splitting probability of reaching the head and binding to the absorbing targets,
\beq\label{eq:split}
P_n(0) = \frac{1}{2 + \frac{L_n}{R_n} + \frac{\pi R_n}{4An_r}} \approx 0.18 \,,
\eeq
and also the corresponding conditional MFPT,
\beq\label{eq:mfpt3}
T_n(0) = \frac{1}{2 + \frac{L_n}{R_n} + \frac{\pi R_n}{4An_r}}\left(\frac{\pi R^3}{3DAn_r}\left(2 + \frac{L_n}{R_n} + \frac{3R_nL_n^2}{8R^3} + \frac{3R_n^2L_n}{4R^3} \right) + \frac{L_n^2}{D}\left(1 + \frac{L_n}{6R_n} + \frac{R_n}{L_n}\right)\right) \approx 0.98 \, {\rm ms}\,.
\eeq
for particles starting at the base of the neck $(Z=0)$. Hence, nearly one out of five particles will reach the head and bind to the targets, for an average diffusion time of one millisecond, of the same order of magnitude as predicted by the formula \eqref{eq:mfpt2}.\\
Finally, it would be valuable to derive higher-order asymptotic formulas revealing the role of the organization of the absorbing windows on the MFPT. Such expansions were developed for the unit ball \cite{cheviakov2010} using strong localized perturbation theory. Another possible extension would be to develop a narrow escape theory for more general composite network domains than the ones considered in Section \S \ref{sec:net}, with different number of connections per each node, narrow passages of various lengths that could not necessarily be oriented perpendicularly to each node, and also with multiple exit sites, potentially located on several spherical compartments.
\section*{Acknowledgements}
F.P.-L.\ was supported by a postdoctoral fellowship from the Fondation ARC (ARCPDF12020020001505). D.H.\ was supported by the European Research Council (ERC) under the European Union’s Horizon 2020 research and innovation program (grant agreement No 882673).
\begin{appendix}
\section{Appendix: Mixed boundary-value problems in the unit ball}\label{sec:head}
In this appendix we restrict our analysis of the composite domain to the spherical section $\omega_h$, defined as the unit ball $\omega_h = \{\x \in \R^3 \, | \, \|\x\| \leq 1 \}$ with its center conveniently shifted to the origin. On the boundary there are $n_r$ narrow circular targets of equal radius $a$ and centered in $\x_j \in \p\omega_h$, defined as
\beq
\p\omega_{hj} = \{\x \in \p\omega_h \, | \, \|\x-\x_j\| \leq a \}\,, \quad j = 1,2\ldots n_r\,.
\eeq
An additional window or radius $r_n$, corresponding to the intersection between the spherical boundary $\p\omega_h$ and the narrow passage $\p\omega_n$, is centered around the South Pole $\x_h$ as
\beq
\p\omega_n\cap\p\omega_h = \{\x \in \p\omega_h \, | \, \|\x-\x_h\| \leq r_n \}\,.
\eeq
We then recall the mixed boundary-value problems satisfied by the splitting probability $p_h(\x)$,
\beq\label{eq:poisson_split}
\begin{split}
&\Delta p_h(\x) = 0\,, \quad \x \in \omega_h\,, \quad p_h(\x) = 1\,, \quad \x \in \cup_{j=1}^{n_r}\p\omega_{hj}\,, \\
&\frac{\p}{\p \n}p_h(\x) = 0\,, \quad \x \in \p\omega_h \backslash \left\{ \cup_{j=1}^{n_r}\p\omega_{hj}\cup\p\omega_{n} \right\}\,, \quad \frac{\p}{\p \n}p_h(\x) = \left.-\frac{d}{dz}p_n(z)\right|_{z=l_n}\,, \quad \x \in \p\omega_n\cap\p\omega_h\,,
\end{split}
\eeq
and by the intermediate variable $v_h(\x)$,
\beq\label{eq:poisson_cond}
\begin{split}
&\Delta v_h(\x) = -p_h(\x)\,, \quad \x \in \omega_h\,, \quad v_h(\x) = 0\,, \quad \x \in \cup_{j=1}^{n_r}\p\omega_{hj}\,, \\
&\frac{\p}{\p \n}v_h(\x) = 0\,, \quad \x \in \p\omega_h \backslash \left\{ \cup_{j=1}^{n_r}\p\omega_{hj}\cup\p\omega_{n} \right\}\,, \quad \frac{\p}{\p \n}v_h(\x) = \left.-\frac{d}{dz}v_n(z)\right|_{z=l_n}\,, \quad \x \in \p\omega_n\cap\p\omega_h\,.
\end{split}
\eeq
Our aim is to derive asymptotic approximations for the splitting probability $p_h(\x_h)$ and the intermediate variable $v_h(\x_h)$ valid in the limit of small, well-separated, narrow passage and targets. That is, we assume $a \ll 1$ and $r_n \ll 1$, as well as $\|\x_h - \x_j\| \sim O(1)$ and $\|\x_i - \x_j\|$ for $j \neq i$. We employ Green's function methods and thus introduce the Neumann Green's function $G_s(\x;\y)$, where the singularity $\y$ is located on the boundary $\p\omega_h$, and which satisfies
\beq\label{eq:GF_pde}
\begin{split}
&\Delta G_s(\x;\y) = \frac{1}{|\omega_h|}\,, \quad \x \in \omega_h\,, \y \in \p\omega_h\,; \\
&\frac{\p}{\p\n}G_s(\x;\y) = \delta(\x-\y)\,, \quad \x\,, \y \in \p\omega_h\,; \quad \int_{\omega_h} G_s(\x;\y)dx = 0\,, \quad \y \in \p\omega_h\,,
\end{split}
\eeq
and that can be expanded as
\beq\label{eq:GF_expan}
G_s(\x;\y) = \frac{1}{2\pi\|\x-\y\|} - \frac{1}{4\pi}\log\left(\|\x-\y\|\right) + O(1)\,, \quad 0 < \|\x-\y\| \ll 1\,,
\eeq
near the singular diagonal $\y=\x$.\\
Finally because of the small radius limit $a \ll 1$, we can approximate the normal derivatives at each target $\p\omega_{hj}$ with the classical Weber's solution \cite{crank1975},
\beq\label{eq:weber}
\frac{\p}{\p \n}p_h(\x) = \frac{B_j}{\sqrt{a^2 - \|\x-\x_j\|^2}}\,, \quad \frac{\p}{\p \n}v_h(\x) = \frac{C_j}{\sqrt{a^2 - \|\x-\x_j\|^2}}\,, \quad \text{for } \|\x-\x_j\| < a \text{ and } \x \in \p\omega_h\,,
\eeq
where $B_j$ and $C_j$ are unknown constants controlling the exit flux.
\paragraph{Splitting probability $p_h(\x)$.} We first apply the divergence theorem to the Poisson's equation \eqref{eq:poisson_split} to obtain the following compatibility condition
\beq\label{eq:compat_split}
0 = \int_{\omega_h} \frac{\p}{\p\x}p_h(\x) d\x = \left.-\frac{d}{dz}p_n(z)\right|_{z=l_n}\pi r_n^2 + \sum_{j=1}^{n_r}\int_{\p\omega_{hj}} \frac{\p}{\p \n}p_h(\x) d\x\,,
\eeq
and then by defining $r = \|\x - \x_j\|$ as the distance from the center of the window $\p\omega_{hj}$, we can compute the integral
\beq
\int_{\p\omega_{hj}} \frac{\p}{\p \n}p_h(\x) d\x = 2\pi B_j\int_0^a \frac{rdr}{\sqrt{a^2 - r^2}} = 2\pi aB_j\,,
\eeq
and thus the compatibility condition \eqref{eq:compat_split} is reduced to the constraint
\beq\label{eq:constraint_split}
\sum_{j=1}^{n_r} B_j = \frac{r_n^2}{2a} \left.\frac{d}{dz}p_n(z)\right|_{z=l_n}\,.
\eeq
\noindent Next, we recall Green's identity
\beq
\int_{\omega_h}\left( G_s(\x;\y)\Delta p_h(\x) - p_h(\x) \Delta G_s(\x;\y) \right)d\x = \int_{\p\omega_h}\left( G_s(\x;\y)\frac{\p}{\p\n}p_h(\x) - p_h(\x)\frac{\p}{\p\n}G_s(\x;\y)\right)d\x\,,
\eeq
and then upon substituting \eqref{eq:poisson_split} and \eqref{eq:GF_pde}, we obtain that for any $y$ on $\p\omega_h$, $p_h(\y)$ satisfies
\beq\label{eq:bdy_1_split}
p_h(\y) = \overline{p_h} - \left.\frac{d}{dz}v_n(z)\right|_{z=l_n}\int_{\p\omega_n\cap\p\omega_h}G_s(\x;\y)d\x + \sum_{j=1}^{n_r} B_j\int_{\p \omega_{hj}}\frac{G_s(\x;\y)}{\sqrt{a^2 - \|\x-\x_j\|^2}}d\x\,,
\eeq
where the average value $\overline{p_h}$ is defined by
\beq
\overline{p_h} = \frac{1}{|\omega_h|}\int_{\omega_h} p_h(\x) d\x\,.
\eeq
By then setting $\y = \x_i$ within \eqref{eq:bdy_1_split} and because $p_h(\x) = 1$ on each absorbing target, we get
\beq
1 = \overline{p_h} - \left.\frac{d}{dz}v_n(z)\right|_{z=l_n}\int_{\p\omega_n\cap\p\omega_h}G_s(\x;\x_i)d\x + \sum_{j=1}^{n_r} B_j \int_{\p \omega_{hj}}\frac{G_s(\x;\x_i)}{\sqrt{a^2 - \|\x-\x_j\|^2}}d\x\,,
\eeq
which is readily approximated by
\beq\label{eq:bdy_2_split}
1 = \overline{p_h} - \left.\frac{d}{dz}v_n(z)\right|_{z=l_n}\pi r_n^2G_s(\x_h;\x_i) + \int_{\p\omega_{hi}} \frac{B_iG_s(\x;\y)}{\sqrt{a^2 - \|\x-\x_i\|^2}}dx + \sum_{\substack{j=1 \\ j \neq i}}^{n_r} B_j 2\pi a G_s(\x_j;\x_i) \,,
\eeq
since the windows are well separated, i.e.\ $\|\x_h - \x_i\| \sim O(1)$ and $\|\x_i - \x_j\| \sim O(1)$ for $i \neq j$.\\
Next, we employ the expansion for the Green's function \eqref{eq:GF_expan} to compute the singular integral over the small patch $\p\Omega_{hi}$ as
\beq\label{eq:int1}
\begin{split}
\int_{\p\omega_{hi}} \frac{G_s(\x;\y)}{\sqrt{a^2 - \|x-x_i\|^2}}dx &= 2\pi\int_0^a \left( \frac{1}{2\pi r} - \frac{1}{4\pi}\log(r) + O(1)\right) \frac{rdr}{\sqrt{a^2 - r^2}}\,, \\
&= \frac{\pi}{2} - \frac{a}{2}\log(a) + O(a)\,,
\end{split}
\eeq
and thus equation \eqref{eq:bdy_2_split} reduces to
\beq\label{eq:bdy_3_split}
1 = \overline{p_h} - \left.\frac{d}{dz}v_n(z)\right|_{z=l_n}\pi r_n^2G_s(\x_h;\x_i) + B_i\left(\frac{\pi}{2} - \frac{a}{2}\log(a) + O(a) \right) + \sum_{\substack{j=1 \\ j \neq i}}^{n_r} B_j 2\pi a G_s(\x_j;\x_i) \,.
\eeq
By then summing \eqref{eq:bdy_3_split} from $i=1$ to $n_r$, substituting the constraint \eqref{eq:constraint_split}, and dropping higher order terms, we obtain
\beq
n_r = n_r\overline{p_h} + \frac{r_n^2}{2a} \left.\frac{d}{dz}p_n(z)\right|_{z=l_n}\left(\frac{\pi}{2} - \frac{a}{2}\log(a)\right) + O(r_n^2)\,,
\eeq
from which we establish that the average splitting probability satisfies
\beq\label{eq:phbar}
\overline{p_h} = 1 - \frac{\pi r_n^2}{4an_r} \left.\frac{d}{dz}p_n(z)\right|_{z=l_n}\left(1 - \frac{a}{\pi}\log(a)\right) + O(r_n^2)\,.
\eeq
\noindent Finally, upon setting $\y=\x_h$ in \eqref{eq:bdy_1_split} we obtain
\beq
p_h(\x_h) = \overline{p_h} - \left.\frac{d}{dz}p_n(z)\right|_{z=l_n} \int_{\p\omega_n\cap\p\omega_h}G_s(\x;\x_h)dx + \sum_{j=1}^{n_r} 2\pi a B_jG_s(\x_j;\x_h) \,,
\eeq
and then computing the surface integral over $\p\omega_h\cap\p\omega_n$ yields
\beq\label{eq:int2}
\begin{split}
\int_{\p\omega_n\cap\p\omega_h}G_s(\x;\x_h)dx &= 2\pi\int_0^{r_n} \left( \frac{1}{2\pi r} - \frac{1}{4\pi}\log(r) + O(1)\right)rdr\,, \\
&= r_n - \frac{r_n^2}{4}\log(r_n) + O(r_n^2)\,,
\end{split}
\eeq
from which we obtain that
\beq\label{eq:phxh}
p_h(\x_h) = 1 - \left.\frac{d}{dz}p_n(z)\right|_{z=l_n}\left(\frac{\pi r_n^2}{4an_r}\left(1 - \frac{a}{\pi}\log(a)\right) + r_n\left(1 - \frac{r_n}{4}\log(r_n) \right) \right) + O(r_n^2)\,,
\eeq
where we used the expression for $\overline{p_h}$ given in \eqref{eq:phbar}.
\paragraph{Intermediate variable $v_h(\x)$.} Similarly as for the splitting probability we first establish the Poisson's compatibility condition of the boundary-value problem \eqref{eq:poisson_cond}, here given by
\beq\label{eq:constraint_cond}
\sum_{j=1}^{n_r} C_j = \frac{1}{2\pi a}\left(\left.\frac{d}{dz}v_n(z)\right|_{z=l_n}\pi r_n^2 - |\omega_h|\overline{p_h}\right)\,.
\eeq
Next, from Green's identity we derive that
\beq\label{eq:bdy_1_cond}
v_h(\y) = \overline{v_h} + \int_{\omega_h}G_s(\x;\y)p_h(\x)d\x - \left.\frac{d}{dz}v_n(z)\right|_{z=l_n}\int_{\p\omega_n\cap\p\omega_h}G_s(\x;\y)dx + \sum_{j=1}^{n_r} C_j\int_{\p \omega_{hj}}\frac{G_s(\x;\y)}{\sqrt{a^2 - \|\x-\x_j\|^2}}d\x\,,
\eeq
where once again the average value $\overline{v_h}$ is defined as
\beq
\overline{v_h} = \frac{1}{|\omega_h|}\int_{\omega_h} v_h(\x) d\x\,.
\eeq
By then neglecting the integral term,
\beq
\int_{\omega_h}G_s(\x;\y)p_h(\x)d\x \approx \int_{\omega_h}G_s(\x;\y)d\x = 0\,,
\eeq
since at leading order the splitting probability equals one, we find that, upon setting $\y = \x_i$ within \eqref{eq:bdy_1_cond} and using the absorbing boundary conditions,
\beq
0 = \overline{v_h} - \left.\frac{d}{dz}v_n(z)\right|_{z=l_n}\int_{\p\omega_n\cap\p\omega_h}G_s(\x;\x_i)d\x + \sum_{j=1}^{n_r} \int_{\p \omega_{hj}}\frac{G_s(\x;\x_i)}{\sqrt{a^2 - \|\x-\x_j\|^2}}d\x\,,
\eeq
which readily approximates as
\beq\label{eq:bdy_2_cond}
0 = \overline{v_h} - \left.\frac{d}{dz}v_n(z)\right|_{z=l_n}\pi r_n^2G_s(\x_h;\x_i) + C_i\left(\frac{\pi}{2} - \frac{a}{2}\log(a) + O(a)\right) + \sum_{\substack{j=1 \\ j \neq i}}^{n_r} C_j 2\pi a G_s(\x_j;\x_i) \,,
\eeq
because the narrow windows are well separated. By next summing \eqref{eq:bdy_2_cond} from $i=1$ to $n_r$, substituting the constraint \eqref{eq:constraint_cond}, and dropping higher order terms we get
\beq
0 = n_r\overline{v_h} + \frac{1}{2\pi a}\left(\pi r_n^2\left.\frac{d}{dz}v_n(z)\right|_{z=l_n} - |\omega_h|\overline{p_h}\right)\left(\frac{\pi}{2} - \frac{a}{2}\log(a) + O(a)\right)\,,
\eeq
from which we can solve for the average value
\beq\label{eq:vhbar}
\overline{v_h} = \frac{1}{4an_r}\left(|\omega_h|\overline{p_h} - \pi r_n^2\left.\frac{d}{dz}v_n(z)\right|_{z=l_n}\right)\left(1 - \frac{a}{\pi}\log(a)\right) + O(a^0)\,.
\eeq
Setting finally $y=\x_h$ within \eqref{eq:bdy_1_cond} yields
\beq
v_h(\x_h) = \overline{v_h} - \left.\frac{d}{dz}v_n(z)\right|_{z=l_n}\int_{\p\omega_n\cap\p\omega_h}G_s(x;\x_h)dx + \sum_{j=1}^{n_r} C_j\int_{\p \omega_{hj}}\frac{G_s(\x;\x_h)}{\sqrt{a^2 - \|\x-\x_j\|^2}}d\x \,,
\eeq
which readily approximates as
\beq
v_h(\x_h) = \overline{v_h} - \left.\frac{d}{dz}v_n(z)\right|_{z=l_n}\left( r_n + \frac{r_n^2}{4}\log(r_n) + O(r_n^2) \right) + \sum_{j=1}^{n_r} 2\pi a C_jG_s(x_j;\x_h) \,,
\eeq
and then by substituting the average value \eqref{eq:vhbar} and dropping higher order terms we get
\beq\label{eq:vhxh}
v_h(\x_h) = \frac{|\omega_h|\overline{p_h}}{4 a n_r}\left(1 - \frac{a}{\pi}\log(a)\right) - \left.\frac{d}{dz}v_n(z)\right|_{z=l_n} \left( \frac{\pi r_n^2}{4 a n_r}\left(1 - \frac{a}{\pi}\log(a)\right) + r_n\left(1 - \frac{r_n}{4}\log(r_n) \right)\right) + O(a^0)\,.
\eeq
\section{Appendix: Boundary layer analysis in the vicinity of the neck}\label{sec:dendrite}
In this Appendix, we proceed to a boundary layer analysis in the dendrite near the intersection with the neck. We consider a cylindrical domain of length $l_d$ and radius $r_d$, parametrized as
\beq
\omega_d = \left\{(x,r,\phi)\, \left| \, -\frac{l_d}{2} \leq x \leq \frac{l_d}{2}\,, \quad 0 \leq r \leq r_d\,, \quad 0 \leq \phi < 2\pi \right. \right\} \,,
\eeq
where $(x,r,\phi)$ are cylindrical polar coordinates defined as
\beq
x = x\,, \quad r = \sqrt{(z+r_d)^2 + y^2}\,, \quad \phi = \arctan\left(\frac{z+r_d}{y}\right)\,.
\eeq
On the boundary $\p\omega_d$ and centered at the origin, there is a narrow circular window of radius $r_n$ and centered at the origin,
\beq
\p\omega_n \cap \p\omega_d = \left\{ (r_d,\phi,x) \in \p\omega_d \, | \, (r_d\phi)^2 + x^2 \leq r_n^2 \right\}\,,
\eeq
that corresponds to the intersection between the dendrite and the neck. The two opposite planar boundaries $\p\omega_{d1}$ and $\p\omega_{d2}$ are defined as
\beq
\p\omega_{dj} = \left\{(x,r,\phi) \in \p\omega_d\, \left| \, x = (-1)^j\frac{l_d}{2}\,, \quad 0 \leq r \leq r_d\,, \quad 0 \leq \phi < 2\pi \right. \right\}\,, \quad j = 1,2\,,
\eeq
\noindent Next, within $\omega_d$ we solve
\beq\label{eq:pde_dendrite}
\Delta p_d(x,r,\phi) = 0\,, \quad \Delta v_d(x,r,\phi) = - p_d(x,r,\phi)\,, \quad \text{for } (r,\phi,x) \in \omega_d\,,
\eeq
where $\Delta$ denotes the Laplacian in cylindrical polar coordinates
\beq
\Delta \equiv \frac{\p^2 }{\p x^2} + \frac{\p^2}{\p r^2} + \frac{1}{r}\frac{\p}{\p r} + \frac{1}{r^2}\frac{\p^2}{\p \phi}\,,
\eeq
subject to Robin boundary conditions on the planar boundary sections,
\beq\label{eq:dend_bc1}
\left[(-1)^j\frac{\p}{\p x} + \lambda\right]p_d(x,r,\phi) = \left[(-1)^j\frac{\p}{\p x} + \lambda\right]v_d(x,r,\phi) = 0\,, \quad \text{for } (x,r,\phi) \in \p\omega_{dj}\,, \quad j=1,2\,,
\eeq
while no-flux boundary conditions are imposed on the curved section,
\beq\label{eq:dend_bc2}
\left.\frac{\p}{\p r}p_d\right|_{r=r_d} = \left.\frac{\p}{\p r}v_d\right|_{r=r_d} = 0\,, \quad \text{for } (x,r_d,\phi) \in \p\omega_d\backslash\p\omega_n\,,
\eeq
except for a narrow circular opening of radius $r_n$ centered at the origin where a Neumann flux condition holds, thereby yielding
\beq\label{eq:dend_bc3}
\left.\frac{\p}{\p r}p_d\right|_{r=r_d} = \left.\frac{d}{dz}p_n\right|_{z=0} \quad \text{and} \quad \left.\frac{\p}{\p r}v_d\right|_{r=r_d} = \left.\frac{d}{dz}v_n\right|_{z=0}\,, \quad \text{for } (x,r_d,\phi) \in \p\omega_d\cap\p\omega_n\,.
\eeq
Here the flux boundary condition comes out of the matching condition with the narrow passage $\omega_n$ that connects the dendrite to the spherical head.
\paragraph{Fully reflective boundary ($\lambda = 0$).} For this case the splitting probability is $p_d \equiv 1$ and a relation for the exit flux through the narrow opening is readily derived using the divergence theorem,
\beq
-|\omega_d| = \int_{\omega_d} \Delta v_d(\x) d\x = \int_{\p\omega_d} \frac{\p}{\p\n}v_d(\x) d\x = \pi r_n^2 \left.\frac{d}{dz}v_n\right|_{z=0}\,,
\eeq
and thus we find
\beq
\left.\frac{d}{dz}v_n\right|_{z=0} = -\frac{|\omega_d|}{\pi r_n^2}\,.
\eeq
\paragraph{Absorbing boundary conditions ($\lambda = \infty$).} We now set $\lambda = \infty$ within \eqref{eq:dend_bc1} and derive a leading order approximation for the splitting probability $p_d(x,r,\phi)$ and the intermediate variable $v_d(x,r,\phi)$ under the following assumption
\beq
r_n \ll 1\,, \quad l_d \sim O(1)\,, \quad r_d \sim O(1)\,.
\eeq
By first neglecting the flux through the narrow window opening we get the outer problem
\beq
\Delta p_d(r,\phi,x) = 0\,, \quad \Delta v_d(r,\phi,x) = - p_d(r,\phi,x)\,,
\eeq
subject to absorbing boundary conditions on the two planar boundaries and no-flux boundary conditions on the curved boundary section,
\beq
\left. p_d\right|_{x = \pm \frac{l_d}{2}} = \left. v_d\right|_{x = \pm \frac{l_d}{2}} = 0\,, \quad \left.\frac{\p}{\p r}p_d\right|_{r=r_d} = \left.\frac{\p}{\p r}v_d\right|_{r=r_d} = 0\,,
\eeq
and thus at leading order we obtain the trivial solution $p_d = v_d = 0$.\\
Next, near the narrow passage $\p\omega_n\cap\p\omega_d$ we parametrize the inner region using a set of local cartesian coordinates given by
\beq\label{eq:cart}
\eta = \frac{r_d - r}{r_n}\,, \quad s_1 = \frac{r_d\phi}{r_n}\,, \quad s_2 = \frac{x}{r_n}\,,
\eeq
and then by defining the inner variable $\mathcal{P}(\eta,s_1,s_2)$ as
\beq
\mathcal{P}(\eta,s_1,s_2) = p_d\left(r_nx,\, r_d - \eta r_n,\, \frac{r_ns_1}{r_d}\right)\,,
\eeq
we find that equation \eqref{eq:pde_dendrite} becomes
\beq\label{eq:inner_pde}
\frac{1}{r_n^2} \left( \frac{\p^2\mathcal{P}}{\p\eta^2} + \frac{\p^2\mathcal{P}}{\p s_1^2} + \frac{\p^2\mathcal{P}}{\p s_2^2} \right) + O\left(\frac{1}{r_n}\right) = 0\,,
\eeq
while the boundary conditions \eqref{eq:dend_bc2} and \eqref{eq:dend_bc3} are reduced to
\beq\label{eq:inner_bc}
\left. -\frac{1}{r_n}\frac{\p \mathcal{P}}{\p \eta}\right|_{\eta = 0} = \left.\frac{d}{dz}p_n\right|_{z=0}\,, \quad s_1^2 + s_2^2 \leq 1\,, \quad \left. -\frac{1}{r_n}\frac{\p \mathcal{P}}{\p \eta}\right|_{\eta = 0} = 0\,, \quad s_1^2 + s_2^2 > 1\,.
\eeq
Away from the inner region the splitting probability vanishes, and thus we impose the far-field condition
\beq
\mathcal{P} \to p_d = 0\,, \quad \text{as} \quad \sqrt{\eta^2 + s_1^2 + s_2^2} \to \infty\,.
\eeq
Upon dropping the error term from \eqref{eq:inner_pde} we get the leading order inner problem, given by
\beq
\frac{\p^2\mathcal{P}}{\p\eta^2} + \frac{\p^2\mathcal{P}}{\p s_1^2} + \frac{\p^2\mathcal{P}}{\p s_2^2} = 0
\eeq
subject to
\beq
\left. -\frac{\p \mathcal{P}}{\p \eta}\right|_{\eta = 0} = r_n \left.\frac{d}{dz}p_n\right|_{z=0} \,, \quad s_1^2 + s_2^2 \leq 1\,, \quad \left. \frac{\p \mathcal{P}}{\p \eta}\right|_{\eta = 0} = 0\,, \quad s_1^2 + s_2^2 > 1\,,
\eeq
and $\mathcal{P} = 0$ at infinity. This corresponds to a steady-state diffusion problem in the infinite half-space, with a heat source supplied  onto the unit disk, and its solution is given in \cite{carslaw1988} as
\beq
\mathcal{P}(\eta,s_1,s_2) = r_n \left.\frac{d}{dz}p_n\right|_{z=0} \int_0^\infty e^{-m\eta}J_0\left(m\sqrt{s_1^2+s_2^2}\right)J_1(m)\frac{dm}{m}\,,
\eeq
where $J_n(m)$ is the usual Bessel function of order $n$.\\
By defining next $\mathcal{V}(\eta,s_1,s_2)$ as the second inner variable, we find that it satisfies
\beq\label{eq:inner_cond}
\frac{1}{r_n^2} \left( \frac{\p^2\mathcal{V}}{\p\eta^2} + \frac{\p^2\mathcal{V}}{\p s_1^2} + \frac{\p^2\mathcal{V}}{\p s_2^2} \right) + O\left(\frac{1}{r_n}\right) = - r_n I \int_0^\infty e^{-m\eta}J_0\left(m\sqrt{s_1^2+s_2^2}\right)J_1(m)\frac{dm}{m}\,,
\eeq
subject to the same boundary conditions as in \eqref{eq:inner_bc},
\beq
\left. -\frac{1}{r_n}\frac{\p \mathcal{V}}{\p \eta}\right|_{\eta = 0} = \left.\frac{d}{dz}v_n\right|_{z=0}\,, \quad s_1^2 + s_2^2 \leq 1\,, \quad \left. -\frac{1}{r_n}\frac{\p \mathcal{V}}{\p \eta}\right|_{\eta = 0} = 0\,, \quad s_1^2 + s_2^2 > 1\,,
\eeq
and far-field conditions $\mathcal{V} \to v_d = 0$ as $\sqrt{\eta^2 + s_1^2 + s_2^2} \to \infty$. Then by multiplying \eqref{eq:inner_cond} with $r_n^2$ and neglecting higher order terms, we obtain the exact same leading order problem as for the splitting probability, which solves as
\beq
\mathcal{V}(\eta,s_1,s_2) = r_n \left.\frac{d}{dz}v_n\right|_{z=0}\int_0^\infty e^{-m\eta}J_0\left(m\sqrt{s_1^2+s_2^2}\right)J_1(m)\frac{dm}{m}\,.
\eeq
\noindent To investigate how solution decay within the dendrite, starting from the center point neck, we set $s_1 = s_2 = 0$ and upon computing the integral
\beq
\int_0^\infty e^{-m\eta}\frac{J_1(m)}{m}dm = \frac{1}{\eta + \sqrt{\eta^2 + 1}}\,,
\eeq
and transforming back to cylindrical polar coordinates, we obtain the asymptotic relations
\beq
p_d(r,0,0) = \left.\frac{d}{dz}p_n\right|_{z=0}\frac{r_n^2}{r_d - r + \sqrt{(r_d - r)^2 + r_n^2}} \quad \text{and} \quad v_d(r,0,0) = \left.\frac{d}{dz}v_n\right|_{z=0}\frac{r_n^2}{r_d - r + \sqrt{(r_d - r)^2 + r_n^2}}\,.
\eeq
On the boundary when $r = r_d$ we therefore get
\beq\label{eq:asymptotic_result_dendrite}
p_d(r_d,0,0) = \left.\frac{d}{dz}p_n\right|_{z=0} r_n \quad \text{and} \quad v_d(r_d,0,0) = \left.\frac{d}{dz}v_n\right|_{z=0} r_n\,,
\eeq
while at the center of the dendrite, when $r=0$,
\beq
p_d(0,0,0) = \left.\frac{d}{dz}p_n\right|_{z=0}\frac{r_n^2}{r_d + \sqrt{r_d^2 + r_n^2}} \quad \text{and} \quad v_d(0,0,0) = \left.\frac{d}{dz}v_n\right|_{z=0} \frac{r_n^2}{r_d + \sqrt{r_d^2 + r_n^2}}\,.
\eeq
which reduces to
\beq
p_d(0,0,0) \approx \left.\frac{d}{dz}p_n\right|_{z=0}\frac{r_n^2}{2r_d} \quad \text{and} \quad v_d(0,0,0) \approx \left.\frac{d}{dz}v_n\right|_{z=0} \frac{r_n^2}{2r_d}\,.
\eeq
because $r_n \ll r_d$.
\section{Appendix: Stochastic simulation algorithm}\label{sec:stoch}
In the results of Fig.\ref{fig:Lambda_inf_fig1} and \ref{fig:Lambda_inf_fig2}, we simulated the diffusion of particles as Brownian motion in a confined geometry shown in Fig.\ref{fig:mito_new}. We implement the Smoluchowski limit of Langevin's equation: $\dot{\X}=\sqrt{2D}\dot{w}$. Here, $w$ is the Wiener white noise delta-correlated in time as well as in space. $\X=\left(X, Y, Z\right)^T$ is the position of a particle at time $t$. This motion of particles is simulated using the Euler's scheme: $\X_n=\X_{n-1}+\sqrt{2 D \Delta t}  \cdot{} \eta$, where $\eta$ is a three-dimensional normal random variable generated by the NumPy library of Python, similar to \cite{basnayak2019Extreme}. The probabilities and binding times are calculated over realisations with 500 particles, with a discrete time-step of $\Delta t = 10^{-7}\, {\rm s}$.
\end{appendix}


\normalem
\bibliographystyle{ieeetr}
\bibliography{bibliomito}

\end{document}